\documentclass[aps,prc,nofootinbib,12pt]{revtex4-2}
\usepackage{graphicx} 

\usepackage{graphicx,epsfig,epstopdf} 
\usepackage{dcolumn}
\usepackage{bm}
\usepackage{amsmath} 

\begin{document}

\title{Discrete and continuous spectrum of lightest hypernuclei}
\author{N. Kalzhigitov}
\email{knurto1@gmail.com}
\affiliation{Al-Farabi Kazakh National University, Almaty, Kazakhstan}

\author{S. Amangeldinova}
\affiliation{Al-Farabi Kazakh National University, Almaty, Kazakhstan}

\author{V. O. Kurmangaliyeva}
\affiliation{Al-Farabi Kazakh National University, Almaty, Kazakhstan}
 
\author{V. S. Vasilevsky}
\email{vsvasilevsky@gmail.com}
\affiliation{Bogolyubov Institute for Theoretical Physics,\\
 Kyiv, 03143, Ukraine
} 


\date{\today}

\begin{abstract}
We analyze the peculiarities of the interaction of the lambda hyperon with s-shell nuclei. The spectra of bound and scattering states are studied in the hypernuclei
$_{\Lambda}^{2}$H, $_{\Lambda}^{3}$H, $_{\Lambda}^{4}$H, $_{\Lambda}^{4}$He and $_{\Lambda}^{5}%
$He, which is considered as two-cluster configurations $p+\Lambda$,
$d$+$\Lambda$, $^{3}$H+$\Lambda$, $^{3}$He+$\Lambda$, $^{4}$He+$\Lambda$, respectively. The
explicit form of the folding potentials of such an interaction is presented in
coordinate and oscillator representations, which help us to understand the
structure of hypernuclei of interest. We compare energies of bound states,
phase shifts of the elastic lambda hyperon, and neutron scattering from s-shell nuclei.

\end{abstract}
\maketitle



\section{Introduction}

Investigations of the lambda hyperon-nucleon interaction and the interaction of the lambda hyperon with atomic nuclei are essential for enhancing our understanding of the structure of ordinary nuclei and hypernuclei, and the dynamics of
interaction of ordinary nuclei with hypernuclei. Our knowledge of the properties of the hyperon-nucleon interactions is based on hypernuclear spectroscopy and hyperon-nucleon scattering data. On this basis, a set of $N\Lambda$ potentials has been proposed, and then these potentials have been used in different theoretical models to study the formation of hypernuclear systems. An increasing amount of experimental data concerning the structure of hypernuclei also stimulates theoretical studies. A large number of experimental laboratories are conducting various types of experimental investigations to uncover new and interesting properties of hypernuclear systems. Systematic experimental information about ground and excited states of light hypernuclei is collected on the website \cite{ChartHyperN2021}. A detailed analysis of experiments performed and planned worldwide is presented in a recently published review 
\cite{2025ChPhL..42j0101C}.

The available experimental findings stimulated our investigation of
light hypernuclei. We aim to study the peculiarities of the interaction
of the lambda hyperon with s-shell nuclei. To provide a deeper understanding of the
subject, we also included an analysis of the interaction of a neutron with the same
nuclei. We are going to make a systematic investigation of bound and scattered
states of light hypernuclei $_{\Lambda}^{2}$H, $_{\Lambda}^{3}$H, $_{\Lambda
}^{4}$H, $_{\Lambda}^{4}$He and $_{\Lambda}^{5}$He. We assume that they are formed by the interaction
of the lambda hyperon with proton, deuteron, $^{3}$H, $^{3}$He  and $^{4}$He nuclei, correspondingly.

This study was also motivated by our desire to study several hypernuclei
within a three-cluster model. Within this model, we are planning to study light
nuclei consisting of lambda hyperon and ordinary nuclei with distinguished
(distinct) two-cluster structure such as $^{6}$Li which has $^{4}$He+d
structure or $^{7}$Li with dominant $^{4}$He+$^{3}$H structure and so on. The
key element of this investigation is two-cluster subsystems formed by the
interaction of the lambda hyperon with the lightest nuclei belonging to the nuclear
s-shell. By analyzing the interaction of the lambda hyperon with the s-shell nuclei, we
obtained some interesting results, which we believe will be interesting to publish.

Let us analyze the experimental information about the bound states of the lightest
nuclei $^{2}$H, $^{3}$H, $^{4}$H, $^{4}$He and $^{5}$He and corresponding hypernuclei
$_{\Lambda}^{2}$H, $_{\Lambda}^{3}$H, $_{\Lambda}^{4}$H, $_{\Lambda}^{4}$He and $_{\Lambda}^{5}%
$He. This information is collected in
Table \ref{Tab:BoundStsExp}. Experimental data for ordinary nuclei are taken from Refs.
\cite{2010NuPhA.848....1P, 1992NuPhA.541....1T,
 2002NuPhA.708....3T}, and experimental data for hypernuclei are
collected on the website \cite{ChartHyperN2021}. Energies of bound states, if they
exist, or resonance states with the dominant decay channels, are displayed in
Table \ref{Tab:BoundStsExp}. One can see that  three nuclei ($^{2}$H, $^{3}$H and $^{4}$He)
have a bound state as the ground state, and two other nuclei are represented
by resonance states. On the other side, four hypernuclei are bound and only
$_{\Lambda}^{2}$H, according to \cite{ChartHyperN2021}, manifests itself as an unbound (resonance) state.%

\begin{table}[ht] \centering
\begin{ruledtabular}  
\caption{Experimental energy $E$, their  total  angular momentum and parity $J^{\pi}$ of bound states of the lightest nuclei and hypernuclei. Energy $E$ and width $\Gamma$ are given in MeV. \label{Tab:BoundStsExp}}%
\begin{tabular}
[c]{ccccccccc}
$^{A}Z$ & $J^{\pi}$ & $E$& $\Gamma$ & Channel & $_{\Lambda}^{A}Z$ &
			$J^{\pi}$ & $E$ & Channel\\\hline
			$^{2}$H & $1^{+}$ & -2.224 & - & $p$+$n$ & $_{\Lambda}^{2}$H & $1^{+}$ &
			4.052 & $p$+$\Lambda$\\
			$^{3}$H & $1/2^{+}$ & -6.257 & - & $d$+$n$ & $_{\Lambda}^{3}$H & $1/2^{+}$ &
			-0.164 & $d$+$\Lambda$\\
			$^{4}$H & $2^{-}$ & 3.190 & 5.42 & $t$+$n$ & $_{\Lambda}^{4}$H & $0^{+}$ &
			-2.169 & $t$+$\Lambda$\\
            $^{4}$He & $0^{+}$ & -20.58 & - & $^{3}$He+$n$ & $_{\Lambda}^{4}$He & $0^{+}$
            & -2.347 & $^{3}$He+$\Lambda$\\
			$^{5}$He & $3/2^{-}$ & 0.798 & 0.648 & $^{4}$He+$n$ & $_{\Lambda}^{5}$He &
			$1/2^{+}$ & -3.102 & $^{4}$He+$\Lambda$\\\hline
\end{tabular}
\end{ruledtabular}  
\end{table}%

Approximately the same objectives were put forward in Ref.
\cite{2002PhRvL..89n2504N}, where ab initio calculations have been performed
for $_{\Lambda}^{3}$H, $_{\Lambda}^{4}$H and $_{\Lambda}^{5}$He with four
lambda-nucleon potentials with the explicit admixture of $\Sigma-N$
interaction. The contributions of different components of the lambda-nucleon potentials
and their role in the formation of bound states of $_{\Lambda}^{3}$H, $_{\Lambda
}^{4}$H and $_{\Lambda}^{5}$He have been studied in detail.

The hypernuclei $_{\Lambda}^{3}$H, $_{\Lambda}^{4}$H, $_{\Lambda}^{4}$He,
$_{\Lambda}^{5}$He and $_{\Lambda}^{7}$He have been studied in Ref.
\cite{2025PhRvL.134g2502L} within the chiral effective field theory. The
no-core shell model was employed to perform calculations of the spectra of these hypernuclei. It was found that in
this model, to reproduce experimental values of bound state energy of the
hypernuclei under consideration, it is necessary to take into account the
hyperon-nucleon-nucleon three-body forces.

To achieve our goals, we employ the resonating group method (RGM), or more
exactly, its the algebraic version, which was formulated in Refs. \cite{kn:Fil_Okhr, kn:Fil81}. Recently, the algebraic version of the RGM was properly adopted to
study the interaction of the lambda hyperon with a system of nucleons. In Ref.
\cite{2021NuPhA101622325N}, it was used to study the structure of the
$_{\Lambda}^{9}$Be hypernucleus, considered as a three-cluster structure
$\alpha+\alpha+\Lambda$, and the hypernucleus $_{\Lambda}^{4}$H, treated as a
three-cluster configuration $d+n+\Lambda$, was studied in Ref. \cite{Nesterov:2021gcp}.

This paper is organized as follows. In Sec. \ref{Sec:Method}, we give a short
outline of the main ideas of the algebraic version of the resonating group and its
differences for application to ordinary nuclei and hypernuclei. Properties of lambda-nucleon and lambda-nucleus potentials are discussed in Sec. \ref{Sec:Potentials}. Section
\ref{Sec:Result} is devoted to the analysis of  
discrete and continuous spectra of the lightest hypernuclei and nuclei in the two-cluster formalism, namely, to the analysis of
energies and wave functions of bound states, phase
shifts of the elastic scattering of the lambda hyperon and neutron on the s-shell
nuclei. The paper is closed with a summary in Sec. \ref{Sec:Conclusions}.

\section{Two-cluster RGM method \label{Sec:Method}}

Here, we present in a short form the main ideas of the resonating group method
which we employ to study two-cluster nuclei and hypernuclei. The wave function
describing the interaction of the lambda hyperon with an s-shell nucleus is sought in
the form%
\begin{equation}
	\Psi_{J}\left(  A+\Lambda\right)  =\left\{  \left[  \Phi_{1}\left(
	A,S_{c}\right)  \Phi_{2}\left(  \Lambda\right)  \right]  _{S}\psi_{L}\left(
	\Lambda,\mathbf{x}\right)  \right\}  _{J}, \label{eq:S001}%
\end{equation}
and the wave function describing the interaction of a neutron with s-shell nuclei has a
similar form%
\begin{equation}
	\Psi_{J}\left(  A+n\right)  =\widehat{\mathcal{A}}\left\{  \left[  \Phi
	_{1}\left(  A,S_{c}\right)  \Phi_{2}\left(  n\right)  \right]  _{S}\psi
	_{L}\left(  n,\mathbf{x}\right)  \right\}  _{J}, \label{eq:S002}%
\end{equation}
where $\Phi_{1}\left(  A,S_{c}\right)  $ is the antisymmetric wave function
describing  the internal motion of $A$ nucleons inside s-shell nuclei, the wave functions
$\Phi_{2}\left(  \Lambda\right)  $ and $\Phi_{2}\left(  n\right)  $ represent
the spin state of the lambda hyperon and the spin-isospin state of the nucleon,
respectively. The square brackets stand for the vector coupling of the spins of
lambda hyperon (neutron) and s-shell nucleus, this coupling creates the total
spin $S$. The curly brackets stand for the vector coupling of the total orbital
momentum $L$ and total spin $S$ into the total angular momentum $J$. One
notices that the main difference between Eqs. (\ref{eq:S001}) and
(\ref{eq:S002}) is the antisymmetrization operator $\widehat{\mathcal{A}}$
which permutes the coordinates of valent neutron and coordinates of nucleons,
comprising a s-shell nucleus, and makes the wave function antisymmetric
$\Psi_{J}\left(  A+n\right)  $ of $A$+1 nucleons.

Within the present model, it is assumed that the wave function $\Phi
_{1}\left(  A,S_{c}\right)  $ is fixed and known, while the wave function
$\psi_{L}\left(  x\right)  $ of a relative motion of lambda hyperon or neutron
and the s-shell nucleus has to be determined by solving the appropriate
Schr\"{o}dinger equation. It is obvious that the wave functions (\ref{eq:S001}%
) and (\ref{eq:S002}) obeys (or more precisely, are approximate solutions to)
the different Schr\"{o}dinger equations:%
\begin{align}
	\left(  \widehat{H}_{An}-E\right)  \Psi_{J}\left(  A+n\right)   &
	=0,\label{eq:S005A}\\
	\left(  \widehat{H}_{AL}-E\right)  \Psi_{J}\left(  A+\Lambda\right)   &  =0,
	\label{eq:S005B}%
\end{align}
where Hamiltonians $\widehat{H}_{An}$ and $\widehat{H}_{AL}$ can be presented
as
\begin{align}
	\widehat{H}_{An}  &  =\widehat{H}_{A}+\widehat{T}_{x}+\sum_{i\in A}\widehat
	{V}_{NN}\left(  \mathbf{r}_{i}-\mathbf{r}_{n}\right)  ,\label{eq:S006A}\\
	\widehat{H}_{An}  &  =\widehat{H}_{A}+\widehat{T}_{x}+\sum_{i\in A}\widehat
	{V}_{N\Lambda}\left(  \mathbf{r}_{i}-\mathbf{r}_{\Lambda}\right)  ,
	\label{eq:S006B}%
\end{align}
and they involve nucleon-nucleon $\widehat{V}_{NN}$ and nucleon-hyperon
$\widehat{V}_{N\Lambda}$ potentials. Besides, both Hamiltonians contain the
Hamiltonian $\widehat{H}_{A}$ describing the internal state of $A$-nucleon
nucleus. The expectation value of this Hamiltonian between wave functions
$\Phi_{1}\left(  A,S_{c}\right)  $ determine the internal energy or binding
energy of the nucleus
\[
\mathcal{E}_{A}=\left\langle \Phi_{1}\left(  A,S_{c}\right)  \left\vert
\widehat{H}_{A}\right\vert \Phi_{1}\left(  A,S_{c}\right)  \right\rangle .
\]
As the wave functions $\Phi_{1}\left(  A,S_{c}\right)  $ is fixed, then the
many-particle Schr\"{o}dinger equations (\ref{eq:S005A}) and (\ref{eq:S005B})
can be reduced to the effective two-body problems, as was shown by J. Wheeler
in Refs. \cite{1937PhRv...52.1083W, 1937PhRv...52.1107W}. To do this, one needs to multiply from the left Eqs. (\ref{eq:S005A}) and (\ref{eq:S005B})
on wave functions $\Phi_{1}\left(  A,S_{c}\right)  $ and to integrate over all
coordinates: spatial, spin and isospin. This procedure leads to a two-body
integro-differential Schr\"{o}dinger equations with nonlocal cluster-cluster
potentials%
\begin{align}
	\int d\widetilde{\mathbf{x}}\left[  \widehat{T}_{x}\delta\left(
	\mathbf{x}-\widetilde{\mathbf{x}}\right)  +\widehat{V}_{n}\left(
	\mathbf{x},\widetilde{\mathbf{x}}\right)  -E\mathcal{N}\left(  \mathbf{x}%
	,\widetilde{\mathbf{x}}\right)  \right]  \psi_{L}\left(  n,\widetilde
	{\mathbf{x}}\right)   &  =0,\label{eq:S008A}\\
	\int d\widetilde{\mathbf{x}}\left[  \left(  \widehat{T}_{x}-E\right)
	\delta\left(  \mathbf{x}-\widetilde{\mathbf{x}}\right)  +\widehat{V}_{\Lambda
	}\left(  \mathbf{x},\widetilde{\mathbf{x}}\right)  \right]  \psi_{L}\left(
	\Lambda,\widetilde{\mathbf{x}}\right)   &  =0, \label{eq:S008B}%
\end{align}
where $\widehat{V}_{n}\left(  \mathbf{x},\widetilde{\mathbf{x}}\right)  $ and
$\widehat{V}_{\Lambda}\left(  \mathbf{x},\widetilde{\mathbf{x}}\right)  $ are
nonlocal potentials, and $\mathcal{N}\left(  \mathbf{x},\widetilde{\mathbf{x}}\right)$ is the norm kernel. By introducing the projection
operators $\widehat{P}_{n}\left(  \mathbf{x}\right)  $ and $\widehat
{P}_{\Lambda}\left(  \mathbf{x}\right)  $%
\begin{align}
	\widehat{P}_{n}\left(  \mathbf{x}\right)   &  =\Phi_{1}\left(  A,S_{c}\right)
	\Phi_{2}\left(  n\right)  \delta\left(  \mathbf{r}_{n}-\mathbf{x}\right)
	,\label{eq:S007A}\\
	\widehat{P}_{\Lambda}\left(  \mathbf{x}\right)   &  =\Phi_{1}\left(
	A,S_{c}\right)  \Phi_{2}\left(  \Lambda\right)  \delta\left(  \mathbf{r}%
	_{\Lambda}-\mathbf{x}\right)  , \label{eq:S007B}%
\end{align}
which allows one to reduce the many-particle problem to a two-body problem, we can
express the nonlocal potentials $\widehat{V}_{n}\left(  \mathbf{x}%
,\widetilde{\mathbf{x}}\right)  $ and $\widehat{V}_{\Lambda}\left(
\mathbf{x},\widetilde{\mathbf{x}}\right)  $ as%
\begin{align}
	&  \widehat{V}_{n}\left(  \mathbf{x},\widetilde{\mathbf{x}}\right)
	=\left\langle \widehat{\mathcal{A}}\widehat{P}_{n}\left(  \mathbf{x}\right)
	\left\vert \sum_{i\in A}\widehat{V}_{NN}\left(  \mathbf{r}_{i}-\mathbf{r}%
	_{n}\right)  \right\vert \widehat{\mathcal{A}}\widehat{P}_{n}\left(
	\widetilde{\mathbf{x}}\right)  \right\rangle \label{eq:S009A}\\
	&  \widehat{V}_{\Lambda}\left(  \mathbf{x},\widetilde{\mathbf{x}}\right)
	=\left\langle \widehat{P}_{\Lambda}\left(  \mathbf{x}\right)  \left\vert
	\sum_{i\in A}\widehat{V}_{NN}\left(  \mathbf{r}_{i}-\mathbf{r}_{\Lambda
	}\right)  \right\vert \widehat{P}_{\Lambda}\left(  \widetilde{\mathbf{x}%
	}\right)  \right\rangle \label{eq:S009B}%
\end{align}
and the norm kernel as
\begin{equation}
	\mathcal{N}\left(  \mathbf{x},\widetilde{\mathbf{x}}\right)  =\left\langle
	\widehat{\mathcal{A}}\widehat{P}_{n}\left(  \mathbf{x}\right)  |\widehat
	{\mathcal{A}}\widehat{P}_{n}\left(  \widetilde{\mathbf{x}}\right)
	\right\rangle . \label{eq:S009C}%
\end{equation}

One immediately notices that the full antisymmetrization in the system of neutrons plus s-shell nuclei leads to appearance of the nonlocal norm kernel in the
two-body Schr\"{o}dinger equation (\ref{eq:S008A}). It is necessary to
underline that the nonlocality in Eqs. (\ref{eq:S008A})  and (\ref{eq:S008B}%
) originates from the antisymmetric form of the wave function (\ref{eq:S002})
and from exchange operators in nucleon-nucleon and nucleon-hyperon potentials.
Both potentials (\ref{eq:S009A}) and (\ref{eq:S009B}) can be represented as a
sum of two components%
\begin{align*}
	\widehat{V}_{n}\left(  \mathbf{x},\widetilde{\mathbf{x}}\right)   &
	=\widehat{V}_{n}^{\left(  F\right)  }\left(  x\right)  \delta\left(
	\mathbf{x}-\widetilde{\mathbf{x}}\right)  +\widehat{V}_{n}^{\left(  r\right)
	}\left(  \mathbf{x},\widetilde{\mathbf{x}}\right)  ,\\
	\widehat{V}_{\Lambda}\left(  \mathbf{x},\widetilde{\mathbf{x}}\right)   &
	=\widehat{V}_{\Lambda}^{\left(  F\right)  }\left(  x\right)  \delta\left(
	\mathbf{x}-\widetilde{\mathbf{x}}\right)  +\widehat{V}_{\Lambda}^{\left(
		r\right)  }\left(  \mathbf{x},\widetilde{\mathbf{x}}\right)  ,
\end{align*}
the first component is the so-called folding or direct potential $\widehat
{V}^{\left(  F\right)  }$, this potential is local and represents the main
part of cluster-cluster interaction. The second, residual component
$\widehat{V}^{\left(  r\right)  }$ is totally nonlocal.

For numerical solutions of Eqs. (\ref{eq:S008A}) and (\ref{eq:S008B}) and for
analysis of the results obtained, we employ the full set of wave functions of the
three-dimensional harmonic oscillator. As they form a complete set of basis
functions, then one can use them to expand wave functions of relative motion
of clusters. This method employs the oscillator wave functions for
representing wave functions of bound and scattering states, is known as the algebraic version of the resonating group method, which was formulated in
Refs. \cite{kn:Fil_Okhr, kn:Fil81}. By expanding the wave function
$\psi_{L}\left(  x\right)  =\psi_{L}\left(  n,x\right)  $  and $\psi
_{L}\left(  x\right)  =\psi_{L}\left(  \Lambda,x\right)  $ over oscillator
functions $\Phi_{nL}\left(  x,b\right)  $ (explicit form of these functions
can be found, for example, in Ref. \cite{2015NuPhA.941..121L})%
\begin{equation}
	\psi_{L}\left(  x\right)  =\sum_{n=0}^{\infty}C_{nL}\Phi_{nL}\left(
	x,b\right)  ,\label{eq:S020}%
\end{equation}
we transform the integro-differential equations (\ref{eq:S008A}) and
(\ref{eq:S008B}) into an infinite system of linear algebraic equations%
\begin{align}
	\sum_{m=0}^{\infty}\left[  \left\langle nL\left\vert \widehat{T}%
	_{x}\right\vert mL\right\rangle -E\left\langle nL|mL\right\rangle
	+\left\langle nL\left\vert \widehat{V}_{n}\right\vert mL\right\rangle \right]
	C_{mL} &  =0,\label{eq:S022A}\\
	\sum_{m=0}^{\infty}\left[  \left\langle nL\left\vert \widehat{T}%
	_{x}\right\vert mL\right\rangle -E\delta_{n,m}+\left\langle nL\left\vert
	\widehat{V}_{\Lambda}\right\vert mL\right\rangle \right]  C_{mL} &
	=0.\label{eq:S022B}%
\end{align}
Here $\left\langle nL\left\vert \widehat{T}_{x}\right\vert mL\right\rangle $,
$\left\langle nL\left\vert \widehat{V}_{n}\right\vert mL\right\rangle
$, $\left\langle nL\left\vert \widehat{V}_{\Lambda}\right\vert mL\right\rangle
$ are matrix elements of the corresponding operators, and $\left\langle
nL|mL\right\rangle $ are matrix elements of the norm kernel. The explicit form
of these matrix elements can be found, for example, in Refs.
\cite{kn:cohstate1E,  kn:cohstate2E}.

We do not dwell on the problem of how to numerically solve these systems of equations
with the finite number of oscillator functions $N_{O}$ to obtain convergent
results for bound and continuous spectrum states, as this problem has been thoroughly discussed in Refs. \cite{kn:Fil_Okhr,
 kn:Fil81, 2015NuPhA.941..121L, 2023UkrJPh..68..3K} .

\section{Interaction of lambda hyperon with s-shell nuclei\label{Sec:Potentials}}

In this section, we analyze properties of the $\Lambda$-nucleon potentials of interaction of the lambda hyperon with the s-shell nuclei. The analysis of the $\Lambda$-nucleus interaction is performed in coordinate and
oscillator representations.

\subsection{Interaction of lambda hyperon with nucleon}

We start with an analytical form of the $N\Lambda$ and $NN$ potentials that
will be used in the present paper. In the present work, 
the nucleon-nucleon
interaction is modelled by 
Hasegawa-Nagata potential (HNP) \cite{potMHN1, potMHN2}, and the interaction of
lambda hyperon with nucleon is modelled by the so-called YNG-NF potential
\cite{1994PThPS.117..361Y}. The central part of $NN$ and $N\Lambda$ potentials
can be presented in the following similar form%
\begin{eqnarray}
	V_{N\Lambda}^{\left(  C\right)  }\left(  \mathbf{r}_{i}-\mathbf{r}_{\Lambda
	}\right)  &=&\sum_{p=E,O}\sum_{S=0,1}\sum_{k=1}^{N_{G}}V_{2S+1,p}^{\left(
		k,N\Lambda\right)  } \label{eq:S010A}\\
        &\times&\exp\left\{  -\frac{\left(  \mathbf{r}_{i}-\mathbf{r}%
		_{\Lambda}\right)  ^{2}}{a_{k}^{2}}\right\}  \widehat{P}_{S}\widehat
	{P}_{p},\\
	V_{NN}^{\left(  C\right)  }\left(  \mathbf{r}_{i}-\mathbf{r}_{j}\right)   
&=&\sum_{p=E,O}\sum_{S=0,1}\sum_{k=1}^{N_{G}}V_{2S+1,p}^{\left(  k,NN\right)
	}\\
    &\times& \exp\left\{  -\frac{\left(  \mathbf{r}_{i}-\mathbf{r}_{j}\right)  ^{2}}%
	{a_{k}^{2}}\right\}  \widehat{P}_{S}\widehat{P}_{p}, \label{eq:S010B}%
\end{eqnarray}
where $\widehat{P}_{S}$ is the projection operator that selects spin $S$ of
interacting particles, and the operator $\widehat{P}_{p}$ selects positive
($p=E$) or negative ($p=O$) parity state of the coordinate part of the two-particle
wave function. One can see that both $NN$ and $N\Lambda$ potentials are a
superposition of several ($N_{G}$) Gaussians. Moreover, each component of the
YNG potential is a function of the Fermi momentum $k_{F}$
\begin{equation}
	V_{2S+1,p}^{\left(  N\Lambda\right)  }\left(  \mathbf{r}\right)  =\sum
	_{i=1}^{3}\left[  a_{i}+b_{i}k_{F}+c_{i}k_{F}^{2}\right]  \exp\left\{
	-\frac{\mathbf{r}^{2}}{a_{k}^{2}}\right\}  , \label{eq:S010C}%
\end{equation}
which is employed as an adjustable parameter.

It is worthwhile noticing that the potential of interaction of a lambda hyperon
with a nucleus is nonlocal due to exchange terms in a $N\Lambda$ potential.
The same is true for the interaction of a neutron with a nucleus, besides, in
this case, the antisymmetrization operator creates additional nonlocality. Therefore, we will analyze the local part of the $NN$ and $N\Lambda$
interactions, which are usually called the direct or folding interaction
(potential). For $NN$ and $N\Lambda$ potentials which have a Gaussian form, the
folding potential can be obtained in a simple analytic form.

It is well known that the resonating group method has one free parameter -
the oscillator length $b$. 
The oscillator length \textit{enters} the many-particle oscillator functions $%
\Phi \left( A_{1},S_{1}\right) $, describing the internal structure of $A_{1}$-nucleon system (see Eqs. (\ref{eq:S001}) and (\ref{eq:S001})), and determines the nucleon density
distribution in s-shell nucleus. It selected very often to minimize
the threshold energy of a two- or three-cluster system. Here, we use the same
approach. However, we did not select  the oscillator length $b$ for each nucleus
(hypernucleus), considered in the present paper. We selected a common
oscillator length for all nuclei and hypernuclei, this length minimizes the
energy of the $^{4}$He+$d$ threshold and for the HNP equals $b$=1.357 fm.
As it was pointed out in the Introduction, the present investigation was
stimulated by the investigation, presented in Ref. \cite{2025arXiv250813702K}, of the hypernucleus $_{\Lambda }^{7}$Li within a three-cluster model. This
hypernucleus was considered as a three-cluster configuration  $^{4}$He$%
+d+\Lambda $ and thus the $^{4}$He$+\Lambda $ and $d+\Lambda $ interactions
and the bound states of $_{\Lambda }^{5}$He and $_{\Lambda }^{3}$H play an
important role in the formation of bound and resonance states in  $_{\Lambda
}^{7}$Li. In the present paper, we use the same value of the oscillator
length to study all selected hypernuclei.

\subsection{Folding potentials}

As it was pointed out above, the potential for interaction of the lambda hyperon with a nucleus or a neutron with the same nucleus is nonlocal. However, this potential
includes the folding potential, which is the main part of the lambada-nucleus and
neutron-nucleus potentials. The analysis of the folding potential may explain the
peculiarities of the interaction of the lambda hyperon with the nucleus. That is why in this Section, we will calculate the lambda-nucleus and also neutron - nucleus
potentials. It is well known \cite{1979PhR....55..183S}, that the folding
potential of the interaction of two clusters, comprised of $A_{1}$ and $A_{2}$
nucleons, generated by a nucleon-nucleon potential $\widehat{V}_{NN}\left(
\mathbf{r}_{i}-\mathbf{r}_{j}\right)  $ is
\begin{equation}
	V_{NN}^{\left(  F\right)  }\left(  \mathbf{x}\right)  =\int d\mathbf{r}%
	_{1}d\mathbf{r}_{2}\rho_{1}\left(  \mathbf{r}_{1}\right)  \rho_{2}\left(
	\mathbf{r}_{2}\right)  \widehat{V}_{NN}\left(  \mathbf{r}_{12}\right)  ,
	\label{eq:S030}%
\end{equation}
where $\rho_{1}\left(  \mathbf{r}_{1}\right)  $ and $\rho_{2}\left(
\mathbf{r}_{2}\right)  $ are the matter density distributions within the first
and the second clusters, respectively. As density distributions are determined in
the own system of center of mass, then vector $\mathbf{r}_{12}$ equals%
\begin{equation}
	\mathbf{r}_{12}=\mathbf{r}_{1}-\mathbf{r}_{2}+\mathbf{x,} \label{eq:S030A}%
\end{equation}
where $\mathbf{x}$ is a vector connecting the centers of mass of two clusters.
By using the Fourier transform for the NN potential%
\begin{equation}
	\widehat{V}_{NN}\left(  \mathbf{r}_{12}\right)  =\int d\mathbf{k}%
	\widehat{\mathcal{V}}_{NN}\left(  \mathbf{k}\right)  \exp\left\{  i\left(
	\mathbf{kr}_{12}\right)  \right\}  , \label{eq:S031}%
\end{equation}
we express the folding potential through form factors%
\begin{equation}
	V_{NN}^{\left(  F\right)  }\left(  \mathbf{x}\right)  =\int d\mathbf{k}%
	\widehat{\mathcal{V}}_{NN}\left(  \mathbf{k}\right)  \exp\left\{  i\left(
	\mathbf{kx}\right)  \right\}  F_{1}\left(  \mathbf{k}\right)  F_{2}\left(
	\mathbf{k}\right)  , \label{eq:S032}%
\end{equation}
where%
\begin{align}
	F_{1}\left(  \mathbf{k}\right)   &  =\int d\mathbf{r}_{1}\rho_{1}\left(
	\mathbf{r}_{1}\right)  \exp\left\{  i\left(  \mathbf{kr}_{1}\right)  \right\}
	,\label{eq:S033A}\\
	F_{2}\left(  \mathbf{k}\right)   &  =\int d\mathbf{r}_{2}\rho_{2}\left(
	\mathbf{r}_{2}\right)  \exp\left\{  -i\left(  \mathbf{kr}_{2}\right)
	\right\}  . \label{eq:S033B}%
\end{align}

One notices that, discussing folding potential calculations, we considered
a general case of a two-cluster system with arbitrary values of $A_{1}$ and
$A_{2}$. Now we have to fix them for the cases in our hands. In what follows, the first cluster is an s-shell nucleus
consisting of $A$ nucleons, and the second cluster is a lambda hyperon for
hypernuclear systems or neutron for ordinary nuclear systems. For lambda
hyperon and neutron, which are structureless particles, the density
distribution $\rho_{2}\left(  \mathbf{r}_{2}\right)  $ is the Dirac
delta-function and thus integration over coordinate $\mathbf{r}_{2}$ leads to
simplified expression%
\begin{eqnarray}
	V_{NN}^{\left(  F\right)  }\left(  \mathbf{x}\right)  &=& \int d\mathbf{r}%
	_{1}d\mathbf{r}_{2}\rho_{1}\left(  \mathbf{r}_{1}\right)  \widehat{V}%
	_{NN}\left(  \mathbf{r}_{1}+\mathbf{x}\right)  \label{eq:S034}\\
    &=& \int d\mathbf{k}%
	\widehat{\mathcal{V}}_{NN}\left(  \mathbf{k}\right)  \exp\left\{  i\left(
	\mathbf{kx}\right)  \right\}  F_{1}\left(  \mathbf{k}\right)  , \nonumber	
\end{eqnarray}

It is easy to calculate the mass or proton form factors for the s-shell nuclei in the
shell model. It has the following form
\begin{equation}
	F_{1}\left(  \mathbf{k}\right)  =\exp\left\{  -\frac{1}{4}\frac{A-1}{A}%
	b^{2}\mathbf{k}^{2}\right\}  , \label{eq:S035}%
\end{equation}
where  $b$ is the oscillator length.

Now we consider the derivation of  the coordinate part of the lambda-nucleus
folding potential, the spin part will be considered later.  As it was
indicated above, the $N\Lambda$ potential, we have chosen, consists of a Gaussian and it allows us to obtain a folding potential in a simple analytic form. To
demonstrate it, we need to calculate the Fourier transform of the Gaussian potential.
It can be verified that
\begin{equation}
	V_{0}\exp\left\{  -\frac{\mathbf{r}^{2}}{a^{2}}\right\}  =V_{0}\frac{a^{3}%
	}{\left(  4\pi\right)  ^{3/2}}\int d\mathbf{k}\exp\left\{  -\frac{1}{4}%
	a^{2}\mathbf{k}^{2}-i\left(  \mathbf{k,r}\right)  \right\}  \label{eq:S037}%
\end{equation}
Thus, we need to calculate the integral%
\begin{eqnarray}
	&& V_{NN}^{\left(  F\right)  }\left(  \mathbf{x}\right)  =V_{0}\frac{a^{3}%
	}{\left(  4\pi\right)  ^{3/2}} \label{eq:S038A}\\
    &\times&\int d\mathbf{k}\exp\left\{  -\frac{1}{4}%
	a^{2}\mathbf{k}^{2}-\frac{A-1}{4A}b^{2}\mathbf{k}^{2}+i\left(  \mathbf{kx}%
	\right)  \right\}  , \nonumber%
\end{eqnarray}
to obtain the folding potential. Integrating over vector $\mathbf{k}$, we obtain%
\begin{equation}
	V_{NN}^{\left(  F\right)  }\left(  \mathbf{x}\right)  =V_{0}\gamma^{-3/2}%
	\exp\left\{  -\frac{1}{\gamma}\frac{\mathbf{x}^{2}}{a^{2}}\right\}  ,
	\label{eq:S038B}%
\end{equation}
where%
\begin{equation}
	\gamma=1+\frac{\left(  A-1\right)  }{A}\frac{b^{2}}{a^{2}}. \label{eq:S039}%
\end{equation}

As we can see, similar to the $N\Lambda$ potential, folding $A\Lambda$ potential also has a Gaussian form, with the depth moderated by the factor $ \gamma^{-3/2}$ and radius increased by the factor $\sqrt{\gamma}$. Note that the folding potential of neutron interaction with s-shell nucleus has the same form (\ref{eq:S038B}). Obviously, the depth $V_{0}$  and radius $a$ of NN interaction are different from $N\Lambda$ interaction.

By taking into account the spin components of $N\Lambda$ interaction, we
arrive at an explicit expression of folding potential for the lambda hyperon
interaction with s-shell nuclei:%
\begin{eqnarray}
	&&V_{N\Lambda}^{\left(  F\right)  }\left(  \mathbf{x}\right)     =\sum
	_{k=1}^{N_{G}}\gamma_{k}^{-3/2}\exp\left\{  -\frac{1}{\gamma_{k}}%
	\frac{\mathbf{x}^{2}}{a_{k}^{2}}\right\} \label{eq:S042}\\
	&  \times&\left\{
	\begin{array}
		[c]{cc}%
		\frac{1}{2}\left(  V_{3E}^{\left(  k\right)  }+V_{3O}^{\left(  k\right)
		}\right)  & p+\Lambda,S=1,\\
		\frac{1}{4}\left(  V_{3E}^{\left(  k\right)  }+V_{3O}^{\left(  k\right)
		}+3V_{1E}^{\left(  k\right)  }+3V_{1O}^{\left(  k\right)  }\right)  &
		d+\Lambda,S=\frac{1}{2},\\
		\frac{1}{4}\left(  5V_{3E}^{\left(  k\right)  }+5V_{3O}^{\left(  k\right)
		}+V_{1E}^{\left(  k\right)  }+V_{1O}^{\left(  k\right)  }\right)  &
		t+\Lambda,S=1,\\
		\frac{3}{4}\left(  V_{3E}^{\left(  k\right)  }+V_{3O}^{\left(  k\right)
		}+V_{1E}^{\left(  k\right)  }+V_{1O}^{\left(  k\right)  }\right)  &
		t+\Lambda,S=0,\\
\frac{1}{4}\left(  5V_{3E}^{\left(  k\right)  }+5V_{3O}^{\left(  k\right)
}+V_{1E}^{\left(  k\right)  }+V_{1O}^{\left(  k\right)  }\right)   &
^{3}\text{He}+\Lambda,S=1,\\
\frac{3}{4}\left(  V_{3E}^{\left(  k\right)  }+V_{3O}^{\left(  k\right)
}+V_{1E}^{\left(  k\right)  }+V_{1O}^{\left(  k\right)  }\right)   &
^{3}\text{He}+\Lambda,S=0,\\
		\frac{1}{2}\left(  3V_{3E}^{\left(  k\right)  }+3V_{3O}^{\left(  k\right)
		}+V_{1E}^{\left(  k\right)  }+V_{1O}^{\left(  k\right)  }\right)  &
		\alpha+\Lambda,S=\frac{1}{2}.
	\end{array}
	\right. \nonumber
\end{eqnarray}
Similar expression for the folding potentials of the interaction of a neutron with
s-shell nuclei:%
\begin{eqnarray}
	&&V_{NN}^{\left(  F\right)  }\left(  \mathbf{x}\right)     =\sum_{k=1}^{N_{G}%
	}\gamma_{k}^{-3/2}\exp\left\{  -\frac{1}{\gamma_{k}}\frac{\mathbf{x}^{2}%
	}{a_{k}^{2}}\right\} \label{eq:S043}\\
	&  \times&\left\{
	\begin{array}
		[c]{cc}%
		\frac{1}{2}\left(  V_{33}^{\left(  k\right)  }+V_{31}^{\left(  k\right)
		}\right)  & p+n,S=1,\\
		\frac{1}{8}\left(  3V_{33}^{\left(  k\right)  }+V_{31}^{\left(  k\right)
		}+9V_{13}^{\left(  k\right)  }+3V_{11}^{\left(  k\right)  }\right)  &
		d+n,S=\frac{1}{2},\\
		\frac{1}{2}\left(  2V_{33}^{\left(  k\right)  }+V_{31}^{\left(  k\right)
		}+V_{13}^{\left(  k\right)  }\right)  & t+n,S=1,\\
		\frac{1}{2}\left(  3V_{33}^{\left(  k\right)  }+2V_{13}^{\left(  k\right)
		}+V_{11}^{\left(  k\right)  }\right)  & t+n,S=0,\\
\frac{1}{4}\left(  7V_{3E}^{\left(  k\right)  }+3V_{3O}^{\left(  k\right)
}+V_{1E}^{\left(  k\right)  }+V_{1O}^{\left(  k\right)  }\right)   &
^{3}\text{He}+\Lambda,S=1,\\
\frac{1}{4}\left(  3V_{3E}^{\left(  k\right)  }+3V_{3O}^{\left(  k\right)
}+5V_{1E}^{\left(  k\right)  }+V_{1O}^{\left(  k\right)  }\right)   &
^{3}\text{He}+\Lambda,S=0,\\
		\frac{1}{4}\left(  9V_{33}^{\left(  k\right)  }+3V_{31}^{\left(  k\right)
		}+3V_{13}^{\left(  k\right)  }+V_{11}^{\left(  k\right)  }\right)  &
		\alpha+n,S=\frac{1}{2}.
	\end{array}
	\right. \nonumber
\end{eqnarray}

In the next section we use formulae (\ref{eq:S042}) and (\ref{eq:S043}) to study
folding potentials of interaction of the lambda hyperon and neutron with the s-shell
nuclei, created by the selected $NN$ and $N\Lambda$ potentials.

\subsection{Visualization of potentials}

\begin{figure}[hptb]
\begin{center}
\includegraphics[width=\textwidth]{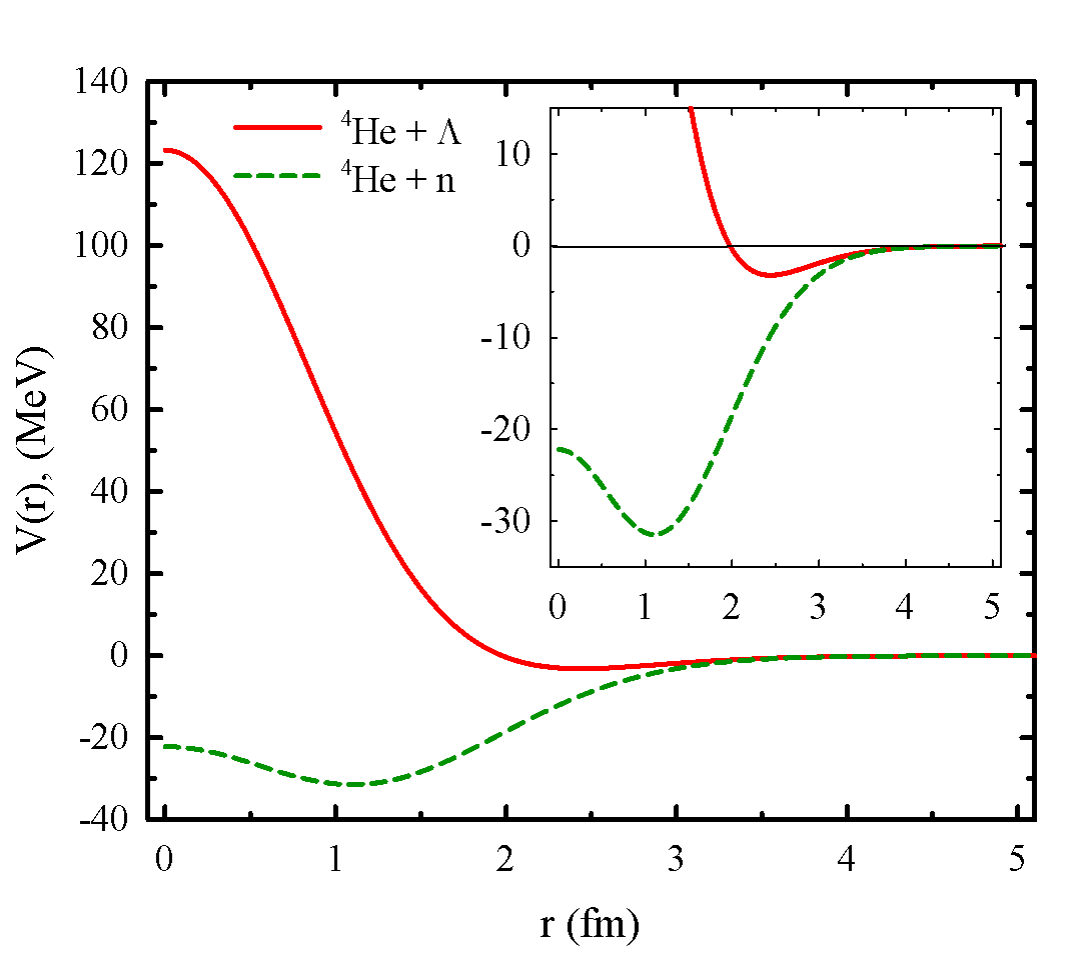}%
\caption{Folding potentials of interaction of the lambda hyperon and neutron with $^{4}$He.}%
\label{Fig:FoldPot5He}%
\end{center}
\end{figure}

In Fig. \ref{Fig:FoldPot5He} we compare folding potentials of interaction of
lambda hyperon and neutron with $^{4}$He. One can see that the $N\Lambda$ folding potential has a large repulsive core and a small attractive well,
while $NN$ folding potential has a rather deep well and a very small repulsive
core. It is necessary to underline that the $^{5}$He nucleus has no bound
state and its ground and excited states are fairly wide resonance states in the state
with the orbital momentum $L$=1 and the total angular momenta $J^{\pi}%
$=3/2$^{-}$ and 1/2$^{-}$, respectively. Contrary to the $^{4}$He$+n$
interaction, the $^{4}$He$+\Lambda$ interaction creates a bound state, as it
will be shown later. 

\begin{figure}[hptb]
\begin{center}
\includegraphics[width=8.6cm]{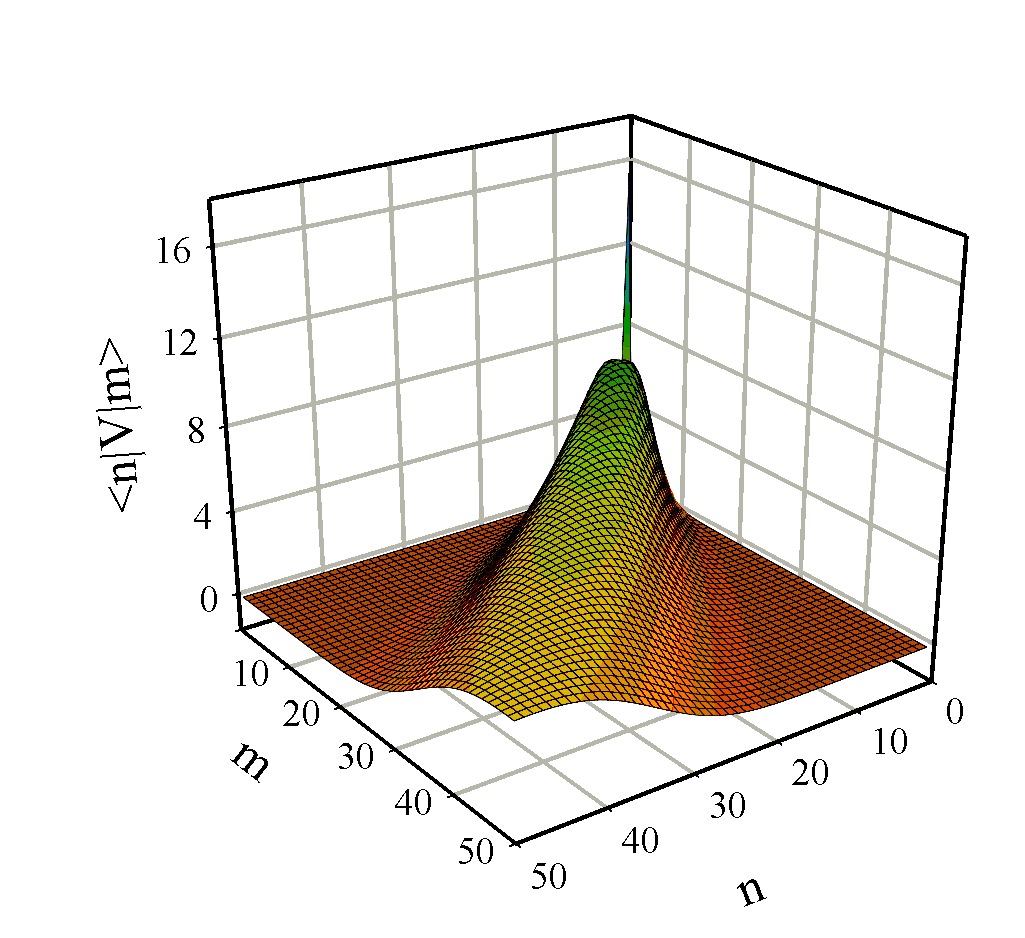}%
\includegraphics[width=8.6cm]{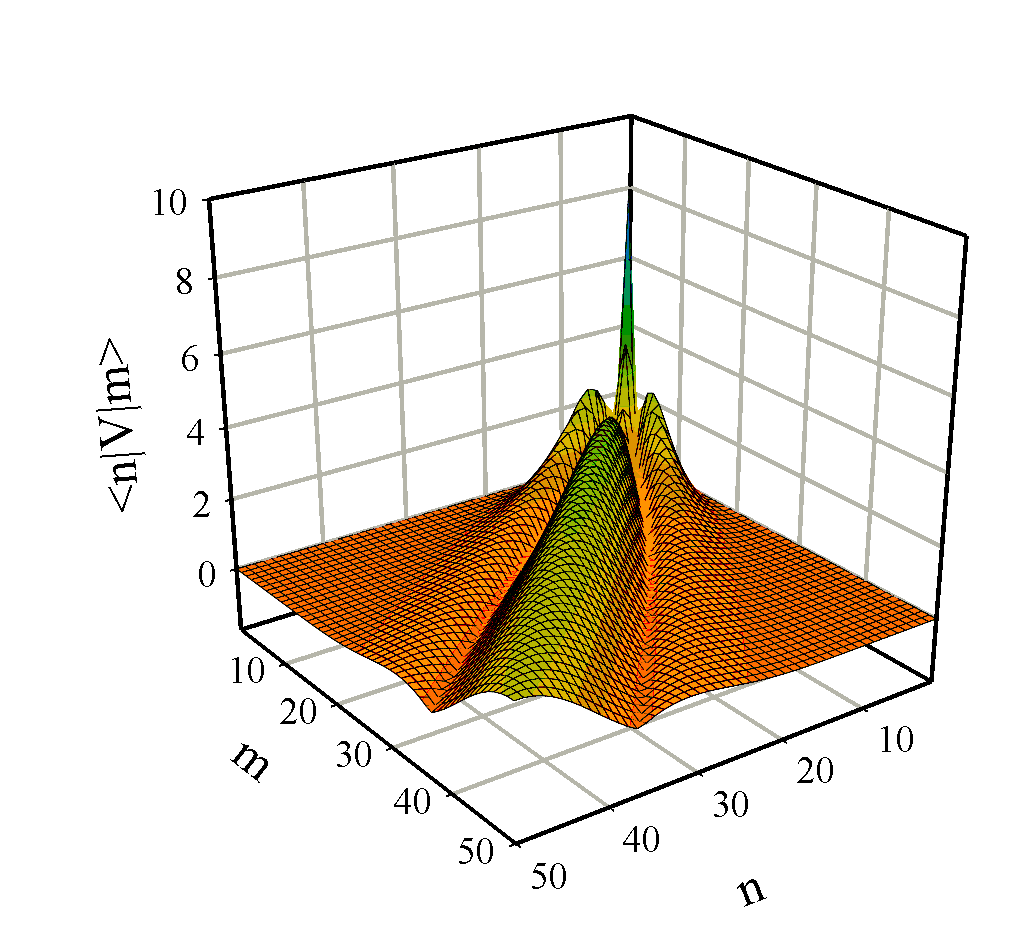}%
\caption{3D presentation of matrix elements of the $\Lambda$+$^{4}$He interaction (top) and $n$+$^{4}$He interaction (bottom).}%
\label{Fig:MEPotEn5He}%
\end{center}
\end{figure}

In Fig. \ref{Fig:MEPotEn5He} and 
we show the 3D picture
of matrix elements of exact $n$+$^{4}$He and $\Lambda
$+$^{4}$He interactions, which involve the exchange operators. This figure
demonstrates that the maximal matrix element of the $\Lambda$+$^{4}$He interaction is approximately three times larger than the maximal matrix element of the $n$+$^{4}$He interaction. It is also seen that all matrix elements of $\Lambda$+$^{4}$He interaction are substantially larger than the
matrix elements of $n$+$^{4}$He interaction. This conclusion is also confirmed
by the next Fig. \ref{Fig:DiagMEPotEn5He}, where diagonal matrix elements of
both interactions are displayed. A long and large tail of diagonal matrix
elements can be attributed to a rather large repulsive core of the $\Lambda$+$^{4}%
$He interaction shown in Fig. \ref{Fig:FoldPot5He}.%

%

\begin{figure}[hptb]
\begin{center}
\includegraphics[width=\textwidth]{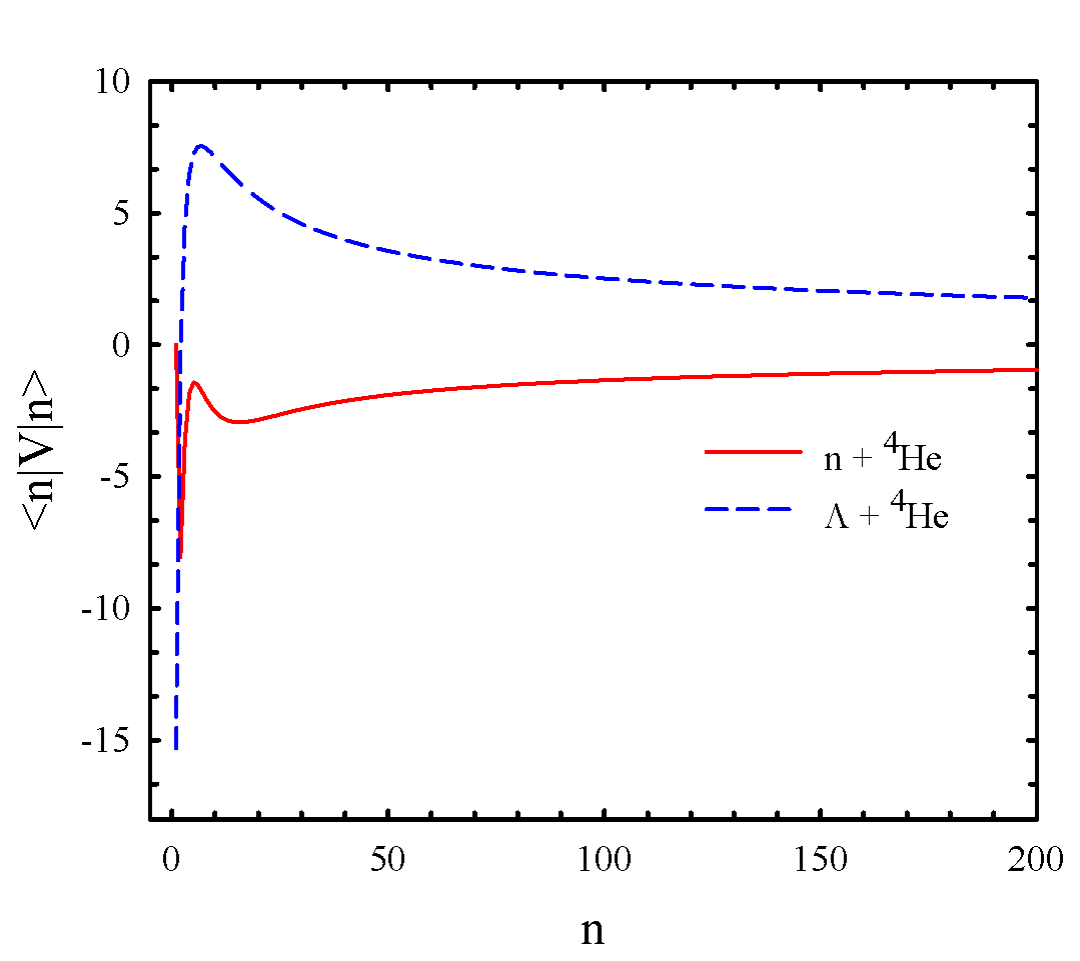}%
\caption{Diagonal matrix elements of potential energy of the $n$+$^{4}$He and $\Lambda$+$^{4}$He interactions.}%
\label{Fig:DiagMEPotEn5He}%
\end{center}
\end{figure}

\begin{figure}[hptb]
\begin{center}
\includegraphics[width=8.6cm]{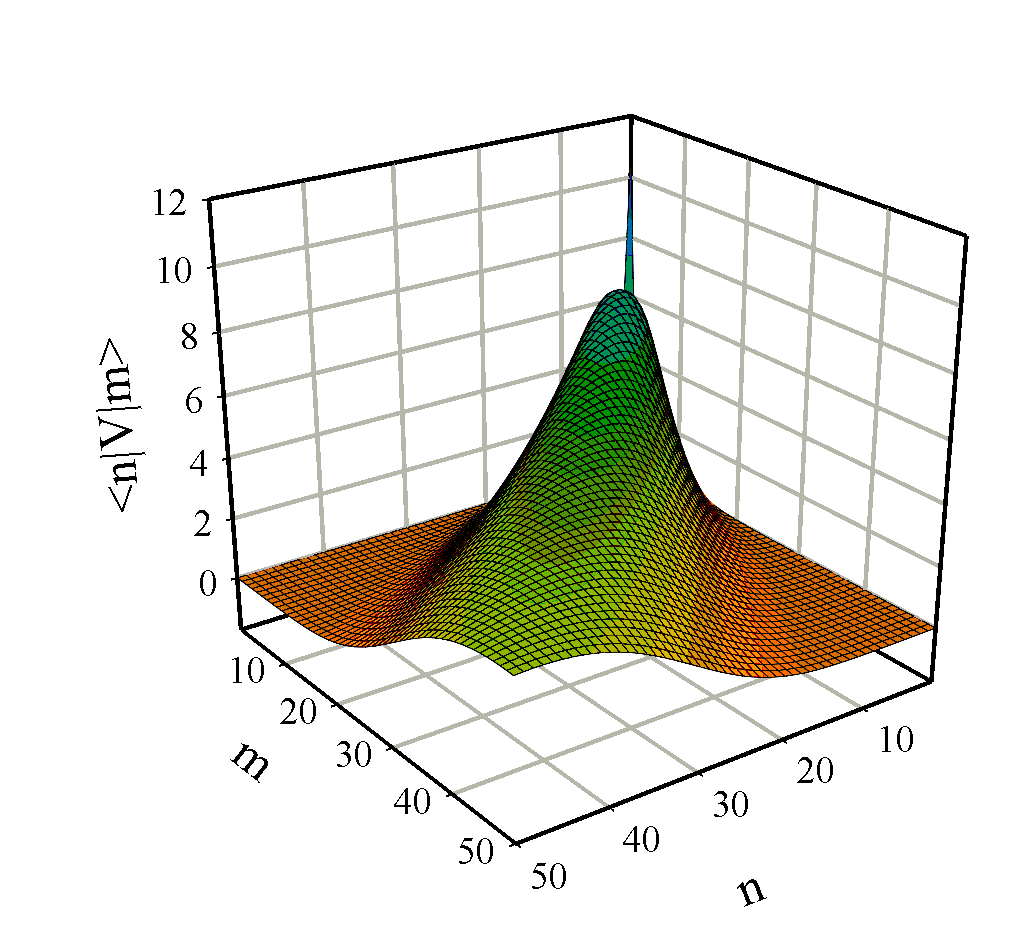}%
\includegraphics[width=8.6cm]{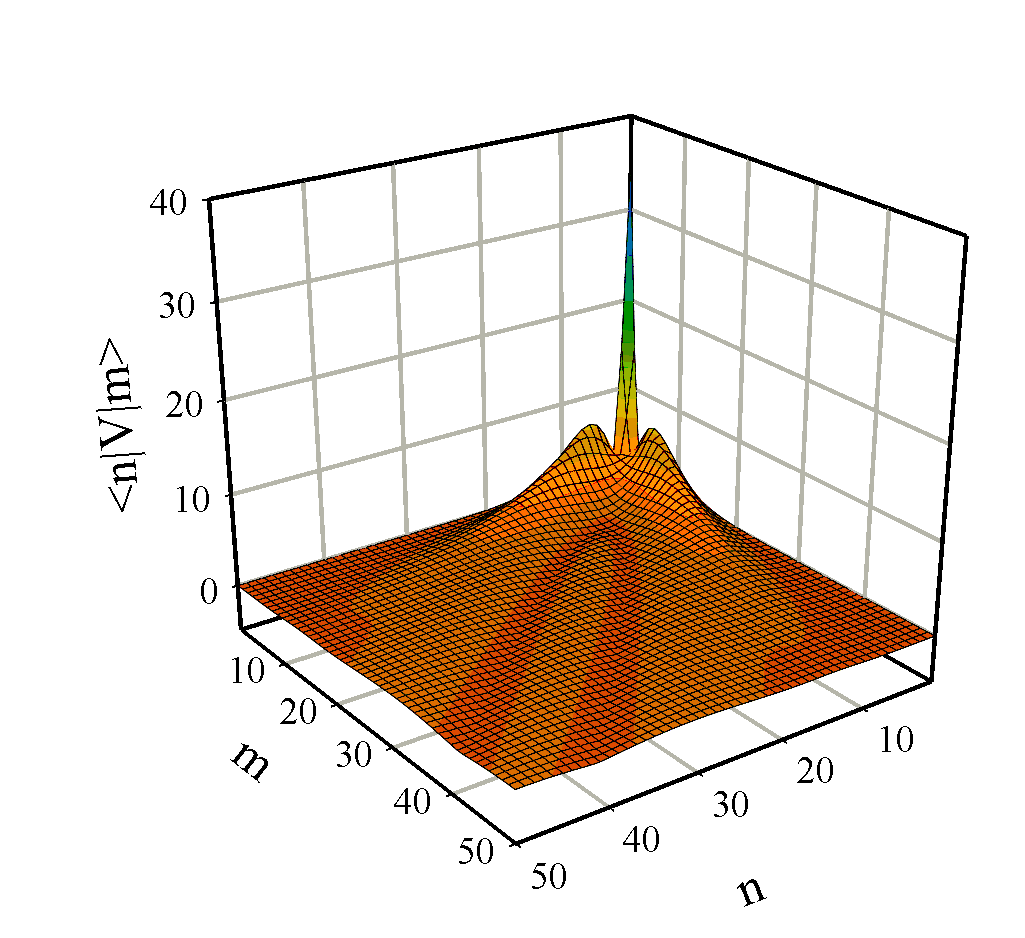}%
\caption{Matrix elements of the potential energy operator for the $\Lambda+d$ interaction (top) and $n+d$ interaction  the  (bottom).}%
\label{Fig:MEPotEn3H3HH}%
\end{center}
\end{figure}

The matrix elements of the potential energy operator for $\Lambda+d$ and $n+d$
interaction are shown in Fig. \ref{Fig:MEPotEn3H3HH}. The matrix elements of the
$n+d$ potential are several times larger than the matrix elements of the
$\Lambda+d$ potential, when $n$ and $m$ are small. Such a difference of matrix
elements explains that the $^{3}$H nucleus has a deeper bound state  than
the hypernucleus $_{\Lambda}^{3}$H.%

\begin{figure}[hptb]
\begin{center}
\includegraphics[width=\textwidth]{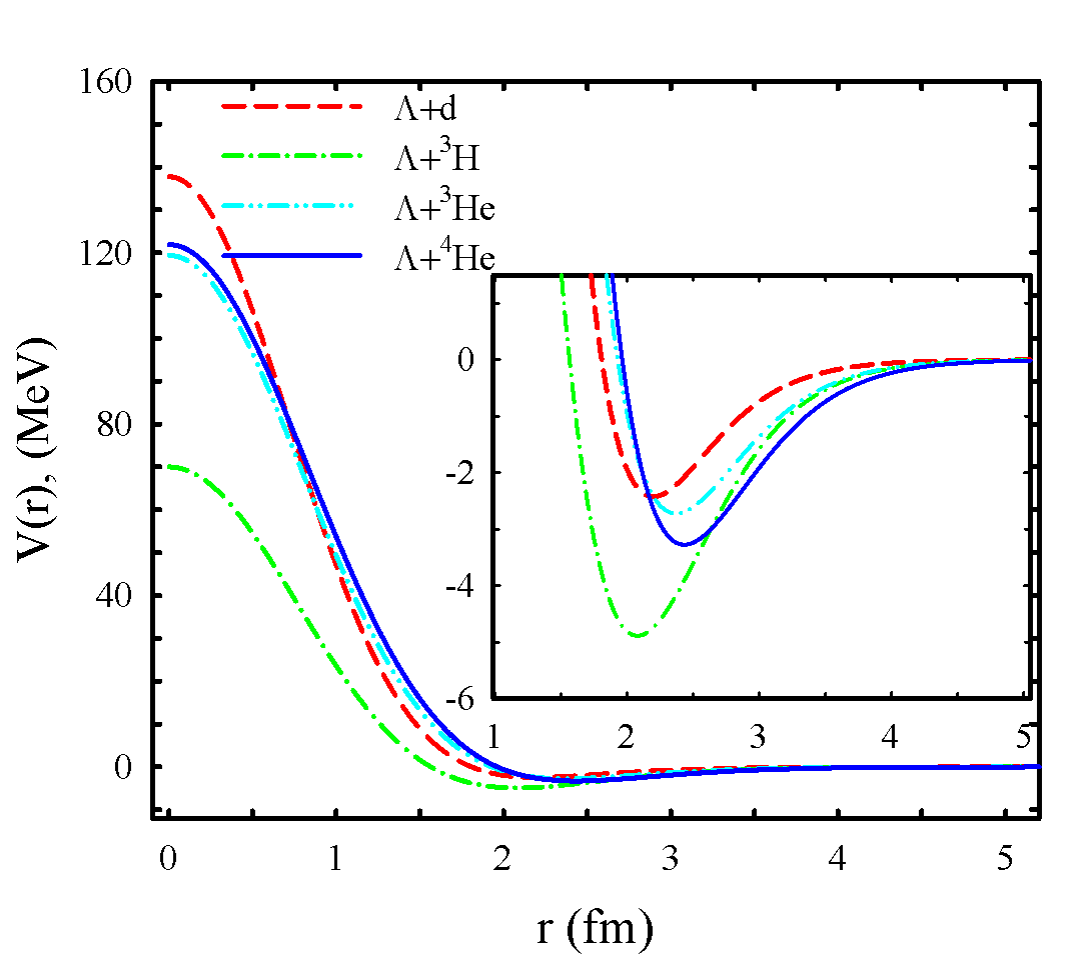}%
\caption{Folding potentials of the $\Lambda$+d, $\Lambda$+$^{3}$H, $\Lambda$+$^{3}$He  and $\Lambda$+$^{4}$He interaction.}%
\label{Fig:FoldPotsBSE}%
\end{center}
\end{figure}

In Fig. \ref{Fig:FoldPotsBSE} we display folding potentials for hypernuclei
$_{\Lambda}^{3}$H, $_{\Lambda}^{4}$H and $_{\Lambda}^{5}$He. The largest
repulsive core is observed in the $d$+$\Lambda$ system and the smallest core is
detected in the $^{3}$H+$\Lambda$ system. The folding potential for $^{3}%
$H+$\Lambda$ system has deepest potential well, and for $d+\Lambda$ it has
smallest potential well. Later, we will demonstrate how it correlates with the
binding energy of these hypernuclear systems.%

\section{Results and discussions\label{Sec:Result}}

\subsection{Bound states properties}

\begin{figure}[hptb]
\begin{center}
\includegraphics[width=\textwidth]{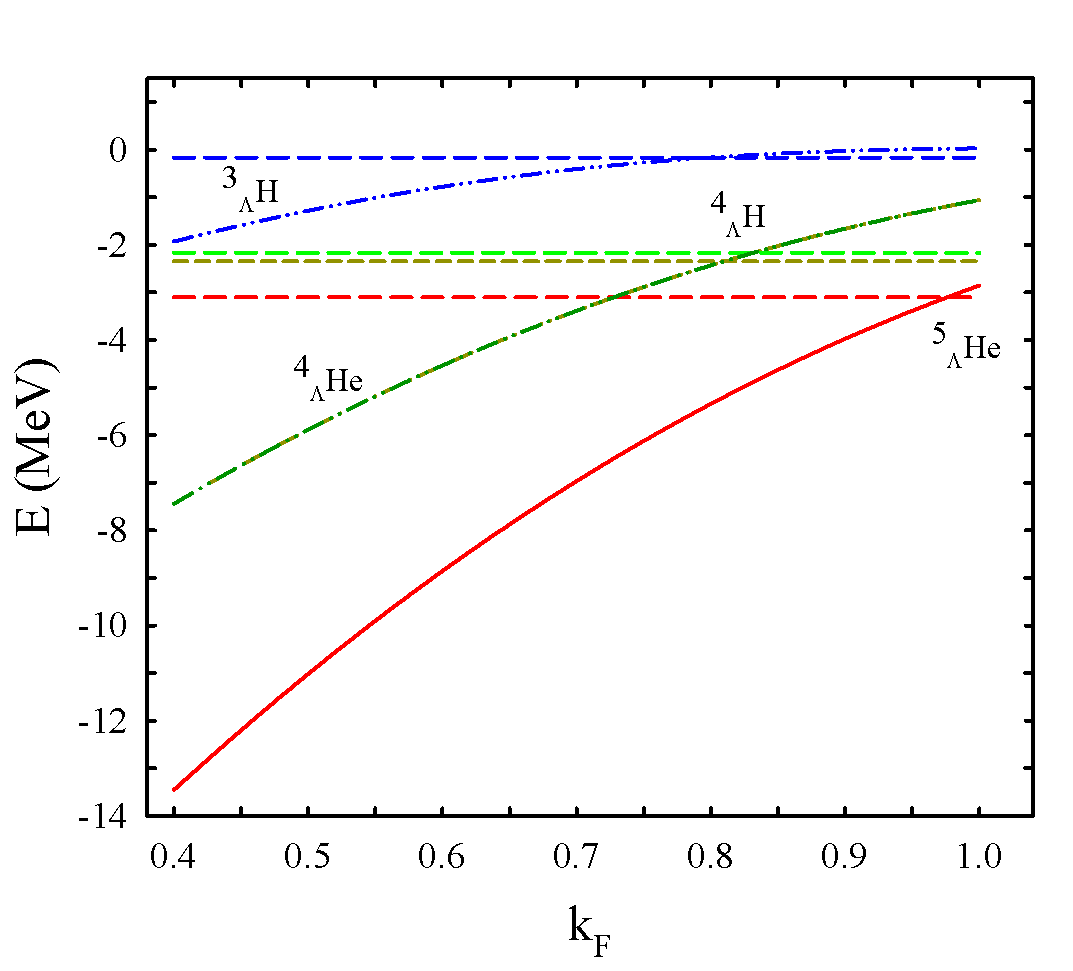}%
\caption{Energy of the ground state of $_{\Lambda}^{3}$H, $_{\Lambda}^{4}$H, $_{\Lambda}^{4}$He  and $_{\Lambda}^{5}$He as a function of the parameter $k_{F}$. The horizontal short-dashed lines indicate the experimental energy of the bound state.}%
\label{Fig:SpectrvsKF}%
\end{center}
\end{figure}

As was pointed out above, the central part of the YNG-NF potential depends on the
parameter $k_{F}$, it usually serves as an adjustable parameter to reproduce,
for example, the ground state energy of a hypernucleus. We will employ this parameter in a similar way. To do this, we consider how the energy of the
ground states of $_{\Lambda}^{3}$H, $_{\Lambda}^{4}$H , $_{\Lambda}^{4}$He and $_{\Lambda}^{5}$He
depends on the parameter $k_{F}$. In Fig. \ref{Fig:SpectrvsKF} we show the
dependence of the binding energy of $_{\Lambda}^{3}$H, $_{\Lambda}^{4}$H, $_{\Lambda}^{4}$He  and
$_{\Lambda}^{5}$He hypernuclei on parameter $k_{F}$. One can see that the
experimental values of the binding energy of these nuclei can be achieved with
different values of the parameter $k_{F}$. One can also see that,  within the present two-cluster model, the ground state energies of $_{\Lambda}^{4}$H and $_{\Lambda}^{4}$He coincide, which reflects the peculiarities of the YNG potential.%

In Table \ref{Tab:EnergyvsKeff} we collected the optimal values of the
parameter $k_{F}$, which allows us to reproduce the ground state energy of
lightest hypernuclei. In what follows, all calculations of discrete and
continuous spectrum states of the hypernuclei $_{\Lambda}^{3}$H, $_{\Lambda
}^{4}$H and $_{\Lambda}^{5}$He will be performed with the corresponding value of
$k_{F}$.%

\begin{table}[ht] \centering
\begin{ruledtabular}  
\caption{Optimal values of $k_{F}$, calculated and experimental energy of ground state $E$ of hypernuclei $_{\Lambda}^{3}$H, $_{\Lambda}^{4}$H, $_{\Lambda}^{4}$He and $_{\Lambda}^{5}$He. \label{Tab:EnergyvsKeff}}%
\begin{tabular}
[c]{cccc}
Hypernucleus & $k_{F}$, fm$^{-1}$ & $E$, MeV & $E_{\exp}$, MeV
			\cite{ChartHyperN2021}\\\hline
			$_{\Lambda}^{3}$H & 0.800 & -0.164 & -0.164\\\hline
			$_{\Lambda}^{4}$H($S$=0) & 0.832 & -2.165 & -2.169\\\hline
            $_{\Lambda}^{4}$He($S$=0) & 0.810 & -2.346 & -2.347\\\hline
			$_{\Lambda}^{5}$He & 0.976 & -3.104 & -3.102\\\hline
\end{tabular}
\end{ruledtabular}  
\end{table}%

\begin{figure}[hptb]
\begin{center}
\includegraphics[width=\textwidth]{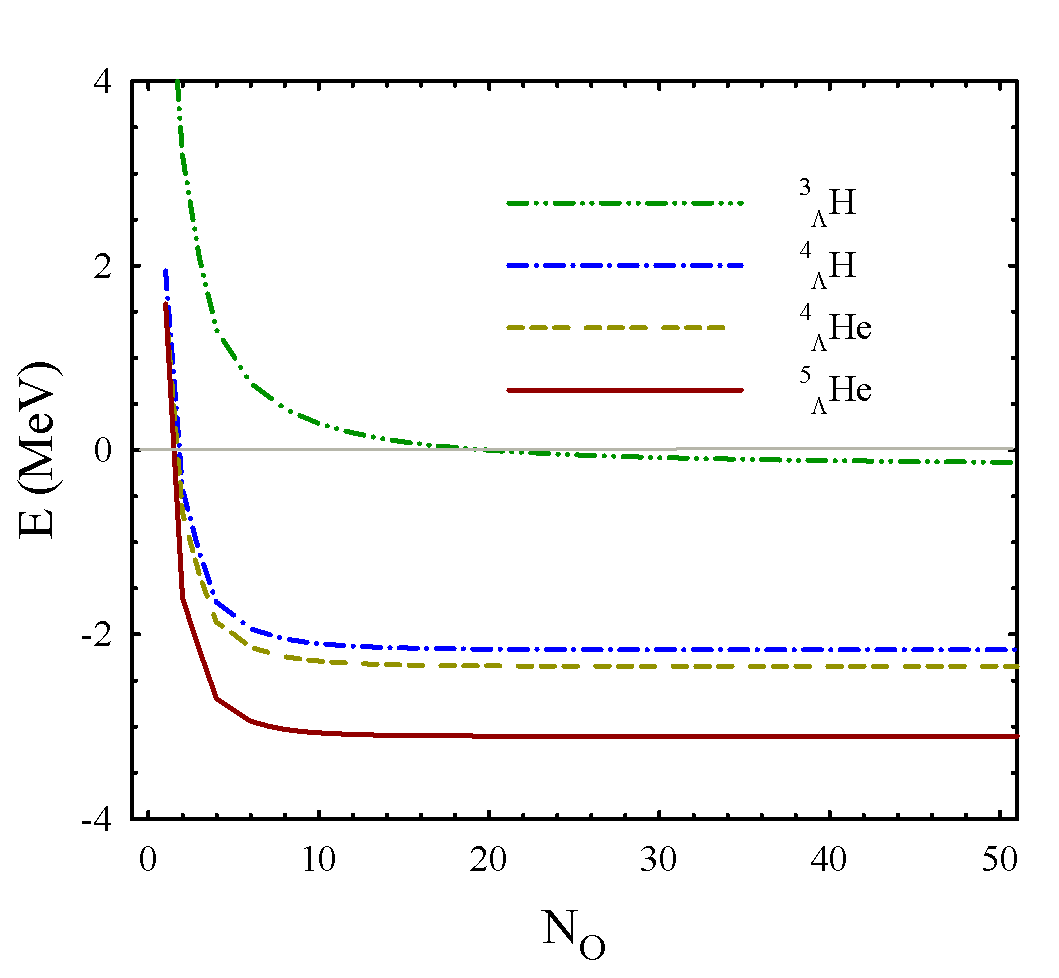}%
\caption{Convergence of the ground state energy of the hypernuclei $_{\Lambda}^{3}$H, $_{\Lambda}^{4}$H, $_{\Lambda}^{4}$He and $_{\Lambda}^{5}$He.}%
\label{Fig:SpectrConvHN}%
\end{center}
\end{figure}

One of the important issues for the present method and for other numerical
methods of solving the many-particle Schr\"{o}dinger equation is the convergence of the
energies and other parameters of the bound states and the scattering parameters for the
continuous spectrum states. Here, we demonstrate the convergence of the ground
state energy of hypernuclei $_{\Lambda}^{3}$H, $_{\Lambda}^{4}$H and $_{\Lambda}^{5}$He. The ground-state energies of these nuclei as a function of
the number of oscillator functions $N_{O}$ are shown in Fig.
\ref{Fig:SpectrConvHN}. One can see that the convergence of the deeply
bound state can be achieved with a relatively small number of oscillator
functions, and the weakly bound state requires a larger number of oscillator
functions. To achieve a precision of 99\% for the ground state of $_{\Lambda}%
^{5}$He (deeply bound state), we need only 11 oscillator functions, and to
achieve the same precision for the ground state of $_{\Lambda}^{3}$H (weakly
bound state), we have to use 160 oscillator functions. In all our calculations of discrete and continuous spectrum states, we employ at least 200 oscillator
functions, which guarantee high precision of the obtained physical quantities.%

In Table \ref{Tab:BoundSts} we collect information about ground states of
hypernuclei of interest. This information includes the ground state energy
$E$, mass root-mean-square radius $R_{m}$ and average distances between
lambda hyperon and s-shell nucleus $A_{c}$. The latter was determined according to the algorithm suggested in Ref. \cite{2023UkrJPh..68..3K}.%

In Table \ref{Tab:BoundSts} we display not only the
ground state of hypernucleus $_{\Lambda }^{4}$H, but also its excited
states.  This state has the total momentum and parity $J^{\pi }$=1$^{+}$,
and in our model it is formed by a zero value of the total orbital momentum $L$=0 and the
total spin $S$=1. For the excited state of  $_{\Lambda }^{4}$H, we used the same
input parameters as for the ground state. For the ordinary nucleus $^{4}$H, we
also calculated both "ground" and "excited"  states, which in fact are
resonance states.

\begin{table}[ht] \centering
\begin{ruledtabular}  
\caption{Parameters of the ground states of lightest hypernuclei and nuclei. Energies of bound and resonance states are given in MeV, mass root-mean-square radii and average distances between clusters are in fm. \label{Tab:BoundSts}}%
\begin{tabular}
[c]{cccccccccc}
$_{\Lambda}^{Z}$A & $J^{\pi}$ & $E$ & $R_{m}$ & $A_{c}$ & $^{Z}$A & $J^{\pi}$
		& $E$ & $R_{m}$ & $A_{c}$\\\hline
		$_{\Lambda}^{3}$H & $\frac{1}{2}^{+}$ & -0.164 & 4.852 & 9.734 & $^{3}$H & $\frac{1}{2}^{+}$ &
		-8.474 & 1.576 & 2.836\\\hline
		$_{\Lambda}^{4}$H($S$=0) & 0$^{+}$ & -2.165 & 1.965 
        & 4.259 & $^{4}$H($S$=0) &
		1$^{-}$ & 1.847 & 9.790 & 22.603\\\hline
		$_{\Lambda}^{4}$H($S$=1) & 1$^{+}$ & -1.623 & 2.111 & 4.576 & $^{4}$H($S$=1) &
		2$^{-}$ & 1.793 & 9.785 & 22.591\\\hline
$_{\Lambda}^{4}$He($S$=0) & 0$^{+}$ & -2.346 & 1.930 & 4.184 & $^{4}$%
He($S$=0) & 0$^{+}$ & -24.082 & 1.433 & 2.358\\\hline
$_{\Lambda}^{4}$He($S$=1) & 1$^{+}$ & -1.787 & 2.061 & 4.469 & $^{4}$%
He($S$=1) & 1$^{+}$ & - & -  & -\\\hline
		$_{\Lambda}^{5}$He & $\frac{1}{2}^{+}$ & -3.104 & 1.828 & 4.271 & $^{5}$He & $\frac{3}{2}^{-}$
		& 0.997 & 6.711 & 16.752\\\hline
\end{tabular}
\end{ruledtabular}  
\end{table}%

As one should expect, the smaller is the absolute value of $E$, the larger is
the mass root-mean-square radius and distance between the lambda hyperon and
s-shell nucleus. The most compact system is the hypernucleus $_{\Lambda}^{5}%
$He with the binding energy $E$=-3.104 MeV and the average distance of $^{4}%
$He+$\Lambda$ system  $A_{c}=$4.27 fm. Comparing the ordinary nucleus $^{3}%
$H (the only nucleus which has a bound state) and hypernucleus $_{\Lambda}%
^{3}$H, we see that the ordinary nucleus is much compact than its counterpart.
The bound state energy of $^{3}$H is approximately five times larger than the
energy of $_{\Lambda}^{3}$H, and the average $d+n$ distance is approximately
3.4 times larger than the $d+\Lambda$ distance.

\subsubsection{Comparing with other models}

In Table \ref{Tab:Spectr4HH3M} we compare our results for the hypernuclei
$_{\Lambda}^{4}$H with the results of two other model calculations. In Ref.
\cite{Nesterov:2021gcp}, a three-cluster model with the three-cluster
configuration $d$+$n$+$\Lambda$ was applied to determine the spectrum of
$_{\Lambda}^{4}$H, in Ref. \cite{2001PhRvC..65a1301H} this hypernucleus was
investigated in a four-cluster model. One can see that our results are fairly close to the results of the three-cluster model \cite{Nesterov:2021gcp}. Note
that our two-cluster model and the three-cluster model \cite{Nesterov:2021gcp}
overbound the first 1$^{+}$ state, while the four-cluster model slightly
underbounds it.%

\begin{table}[ht] \centering
\begin{ruledtabular}  
\caption{Spectrum of bound $^4_{\Lambda}$H states determined in two-, three- and  four-cluster models, which are denoted as 2C, 3C and 4C, respectively. Energy is determined from the $^3$H+$\Lambda$ threshold.\label{Tab:Spectr4HH3M}}%
\begin{tabular}
[c]{ccccc}
& Present 2C  & 3C \cite{Nesterov:2021gcp} & 4C 
		\cite{2001PhRvC..65a1301H} & Exp\\\hline
		$J^{\pi}$ & $E$, MeV & $E$, MeV & $E$, MeV & $E$, MeV\\\hline
		0$^{+}$ & -2.165 & -2.38 & -2.33 & -2.169\\\hline
		1$^{+}$ & -1.623 & -1.56 & -0.59 & -1.081\\\hline
\end{tabular}
\end{ruledtabular}  
\end{table}%

\subsection{Wave functions of bound states}

In this section, we display wave functions of bound states in coordinate and oscillator representations.

\begin{figure}[hptb]
\begin{center}
\includegraphics[width=\textwidth]{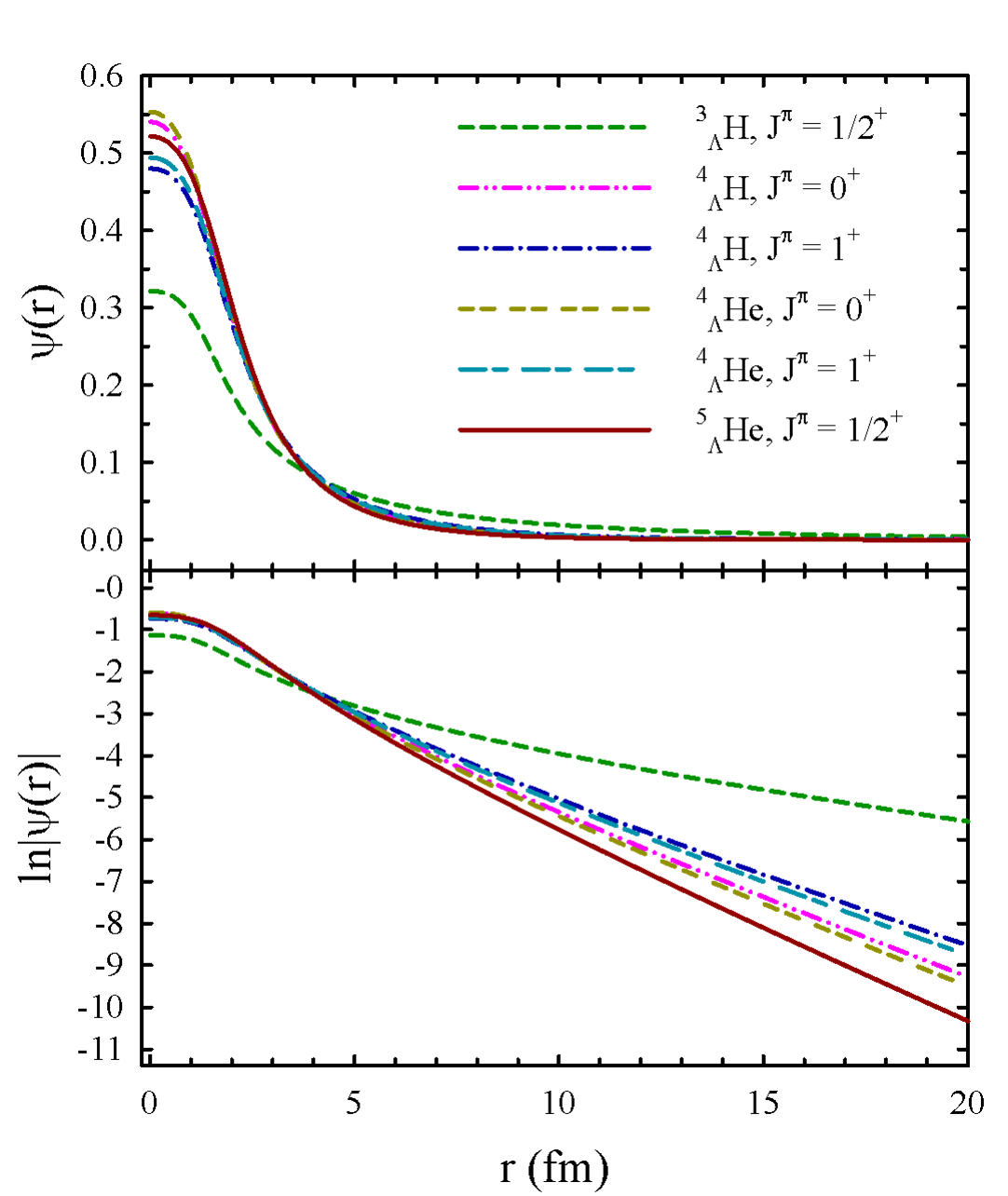}%
\caption{Wave functions of the bound states in $_{\Lambda}^{3}$H, $_{\Lambda}^{4}$H, $_{\Lambda}^{4}$He and $_{\Lambda}^{5}$He in coordinate space.}%
\label{Fig:WaveFuns3LH4LH5LHeCS}%
\end{center}
\end{figure}

In Fig. \ref{Fig:WaveFuns3LH4LH5LHeCS} the wave functions of the bound states
in the $_{\Lambda}^{3}$H, $_{\Lambda}^{4}$H , $_{\Lambda}^{4}$He and $_{\Lambda}^{5}$He as a
function of distance $r$ between lambda hyperon and s-shell nucleus are shown
in normal and logarithmic scales. The  hypernucleus $_{\Lambda}^{4}$H  is
presented by two bound states, one of which is the ground state with quantum
numbers $J^{\pi}$=$0^{+}$, $L=S=0$ and the other is the excited state with
quantum numbers $J^{\pi}$=$1^{+}$, $L=1$, $S=1$. One can see that all
functions have a maximum at $r$=0 fm. The figure in logarithmic scale explicitly demonstrates the exponential behavior of the wave functions at
large distances, and that the hypernucleus $_{\Lambda}^{3}$H has the smallest binding energy and thus the longest tail, and the hypernucleus $_{\Lambda}^{5}$He has the largest binding energy and rapidly decreasing tail.

\begin{figure}[hptb]
\begin{center}
\includegraphics[width=\textwidth]{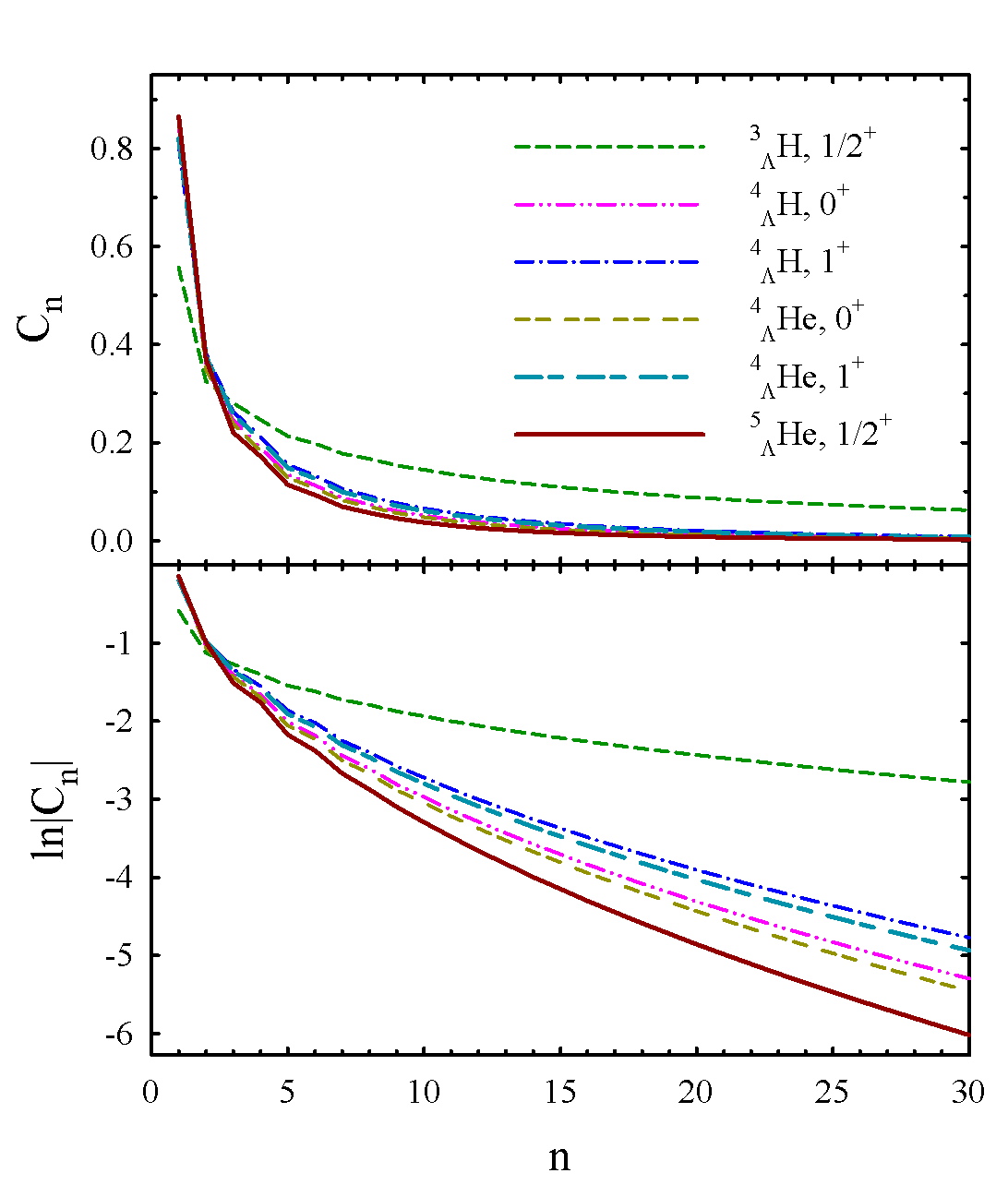}%
\caption{Wave functions $C_{n}$ of bound states of $_{\Lambda}^{3}$H, $_{\Lambda}^{4}$H, $_{\Lambda}^{4}$He and $_{\Lambda}^{5}$He in oscillator representation.}%
\label{Fig:WaveFuns3LH4LH5LHeOS}%
\end{center}
\end{figure}

The behavior of the wave functions $C_{n}$ of the bound states in the oscillator representation displayed in Fig \ref{Fig:WaveFuns3LH4LH5LHeOS} is similar to what is observed in the coordinate representation. Note that for the sake of convenience, these functions are represented by a continuous line, but the actual values of the wave functions are at integer values of $n$. In both representations, wave functions describe a rather compact structure where small distances $r$ and small values of $n$ are dominant. In the asymptotic region of the coordinate (large values of $r$) and oscillator (large values of $n$) spaces, wave functions have an exponential behavior.%

It is worthwhile noticing that the small values of $n$ correspond to small
distances between the clusters and large values of $n$ correspond to large values of
$r$. In Refs. \cite{kn:Fil_Okhr, kn:Fil81}, it was established that the
expansion coefficients $C_{n}$ and the wave function $\psi\left(  r\right)  $ are
related as%
\[
C_{n}=\sqrt{2r_{n}}\psi\left(  r_{n}\right)  ,
\]
where
\[
r_{n}=b\sqrt{4n+2L+3}.
\]
Explicit correspondence between coordinate and oscillator functions has been
demonstrated in Refs. \cite{KALZIGITOV2020Bul, DUISENBAY2019Bul,
2023UkrJPh..68..3K} for the bound and resonance states.

\subsection{Phase shifts}

\begin{figure}[hptb]
\begin{center}
\includegraphics[width=\textwidth]{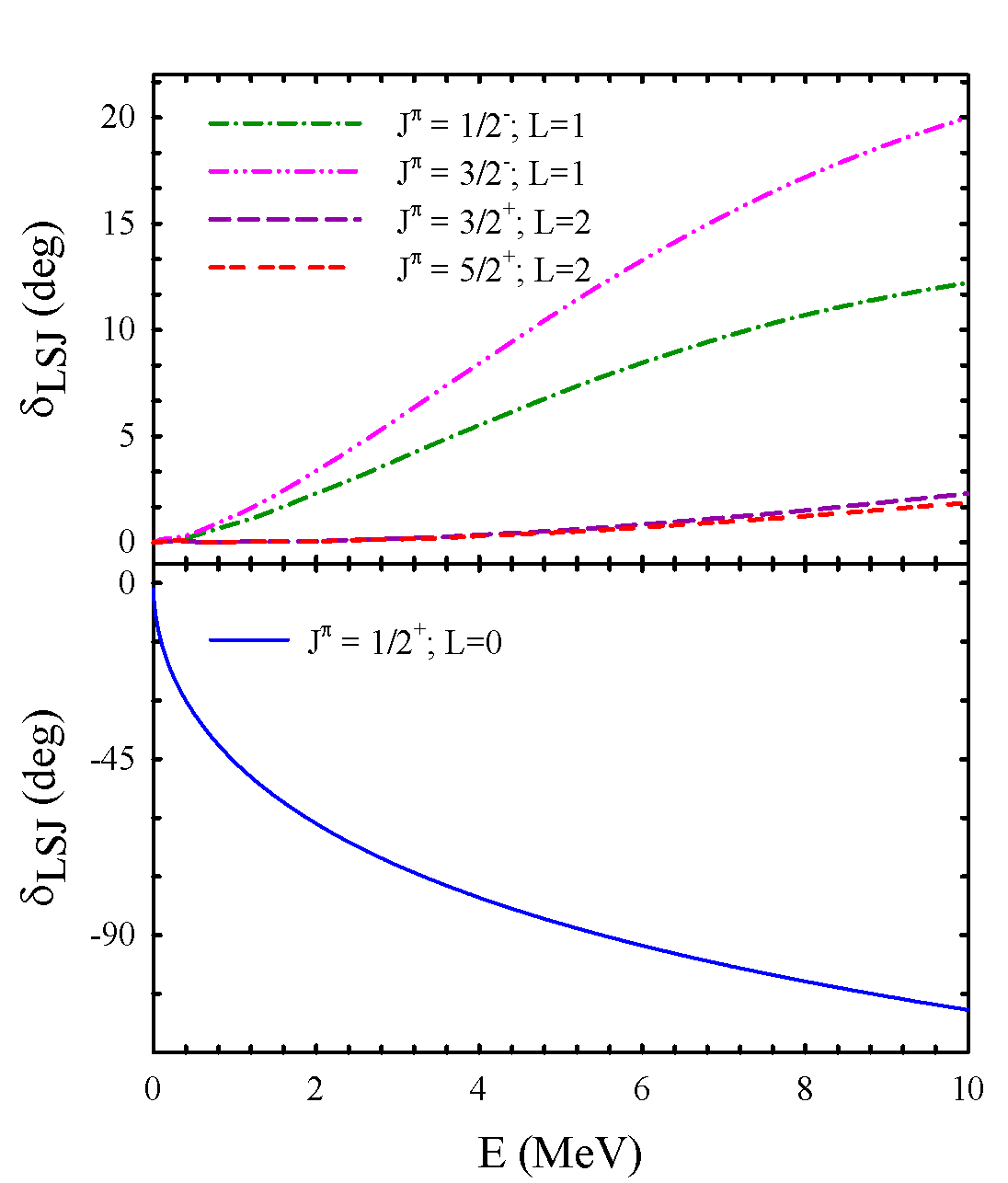}%
\caption{Phase shifts of the elastic scattering of lambda hyperon on $^{4}$He.}%
\label{Fig:PhasesL4He}%
\end{center}
\end{figure}

We calculated phase shifts for elastic scattering of the lambda hyperon and
neutron on the s-shell nuclei. We start the analysis with the phase shifts of
$\Lambda$+$^{4}$He scattering, they are shown in Fig. \ref{Fig:PhasesL4He}.
Phase shifts with the zero orbital momentum ($J^{\pi}$=1/2$^{+}$) are
rapidly decreasing within the selected energy range, while phase shifts with
the orbital momenta $L$=1 and $L$=2 are slowly increasing up to 20$^{\circ}$
($L$=1, $J^{\pi}$=3/2$^{-}$). This behavior of phase shifts reflects the
presence of a bound state below the threshold and that the strongest 
interaction of the lambda hyperon with $^{4}$He is observed in the state with
$L$=0, where a bound state is generated. A shallow potential well with a
combination of the centrifugal barrier does not allow us to create resonance
states in $_{\Lambda}^{5}$He with orbital momenta $L$=1 and $L$=2.

\begin{figure}[hptb]
\begin{center}
\includegraphics[width=\textwidth]{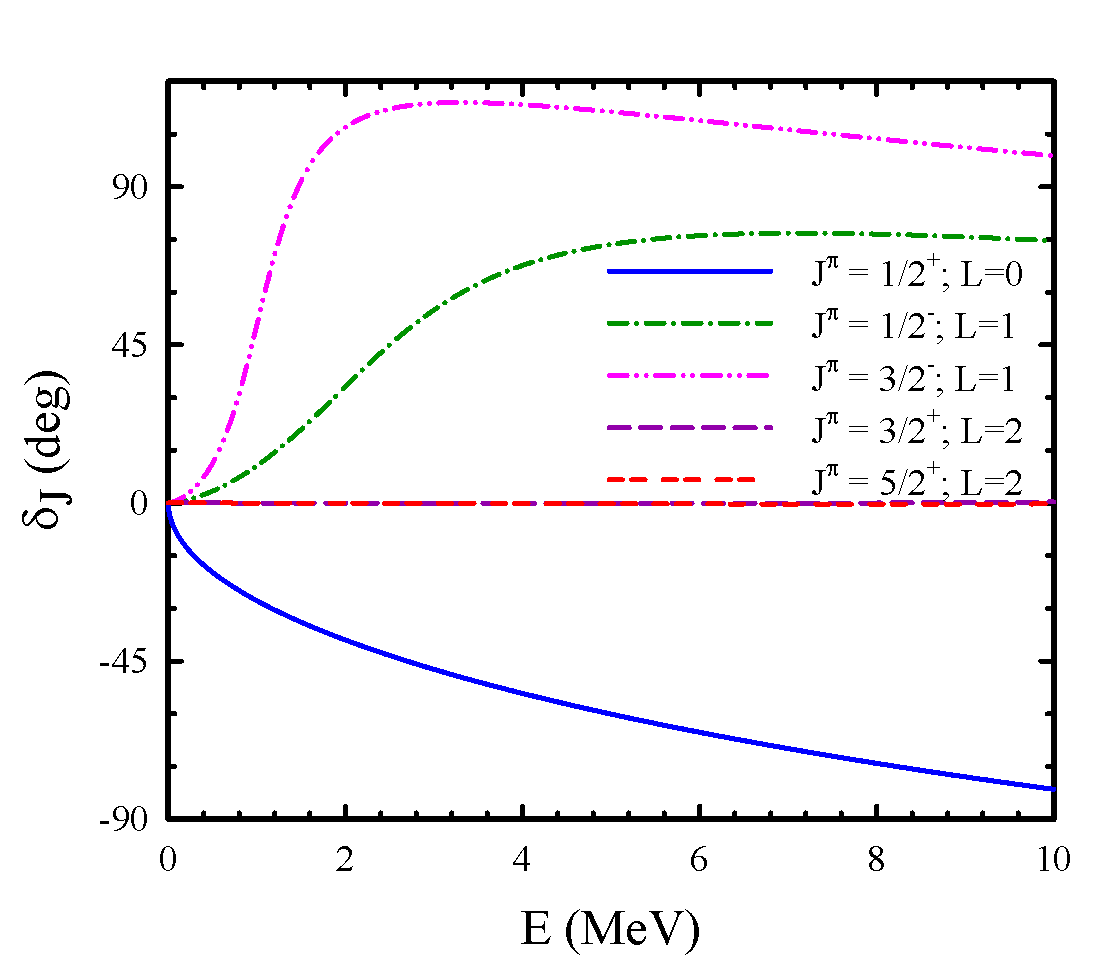}%
\caption{Phase shifts of the elastic $n$+$^{4}$He scattering.}%
\label{Fig:PhasesN4He}%
\end{center}
\end{figure}

Phase shifts of the elastic scattering of neutron on $^{4}$He in the state of positive and negative parities are shown in Fig. \ref{Fig:PhasesN4He}. The phase shifts of the 3/2$^{-}$ and 1/2$^{-}$ states reveal resonance behavior, growing rather fast in small energy regions. The 1/2$^{+}$ phase shift decreases rapidly with increasing energy, indicating that there is fairly strong repulsion between the neutron and $^{4}$He in this state. The behaviour of the 1/2$^{+}$ phase shift of $n$+$^4$He scattering is similar to the behaviour of the phase shift of $\Lambda$+$^4$He scattering. 

\begin{figure}[hptb]
\begin{center}
\includegraphics[width=\textwidth]{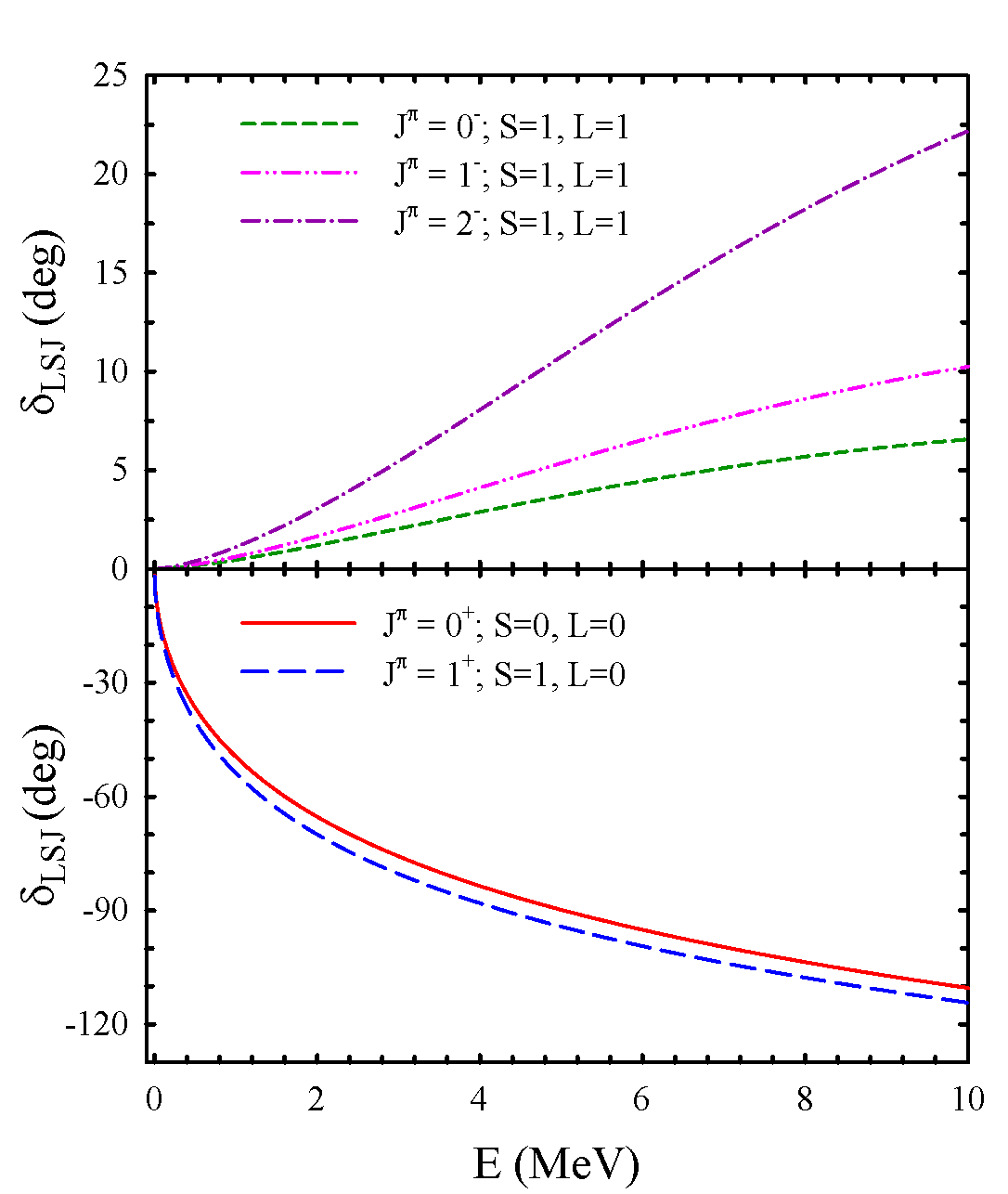}%
\caption{Phase shifts of the $\Lambda$ elastic scattering from $^{3}$H.}%
\label{Fig:PhasesL3H}%
\end{center}
\end{figure}

In Fig. \ref{Fig:PhasesL3H} we demonstrate phase shifts of the elastic
$\Lambda$+$^{3}$H scattering.  One can see that, as in the case of $\Lambda
$+$^{4}$He scattering, the phase shifts of the s-wave of $\Lambda$+$^{3}$H
scattering in the states with the total spin $S$=0 and $S$=1 are rapidly
decreasing and other phase shifts are slowly increasing. %
 
\begin{figure}[hptb]
\begin{center}
\includegraphics[width=\textwidth]{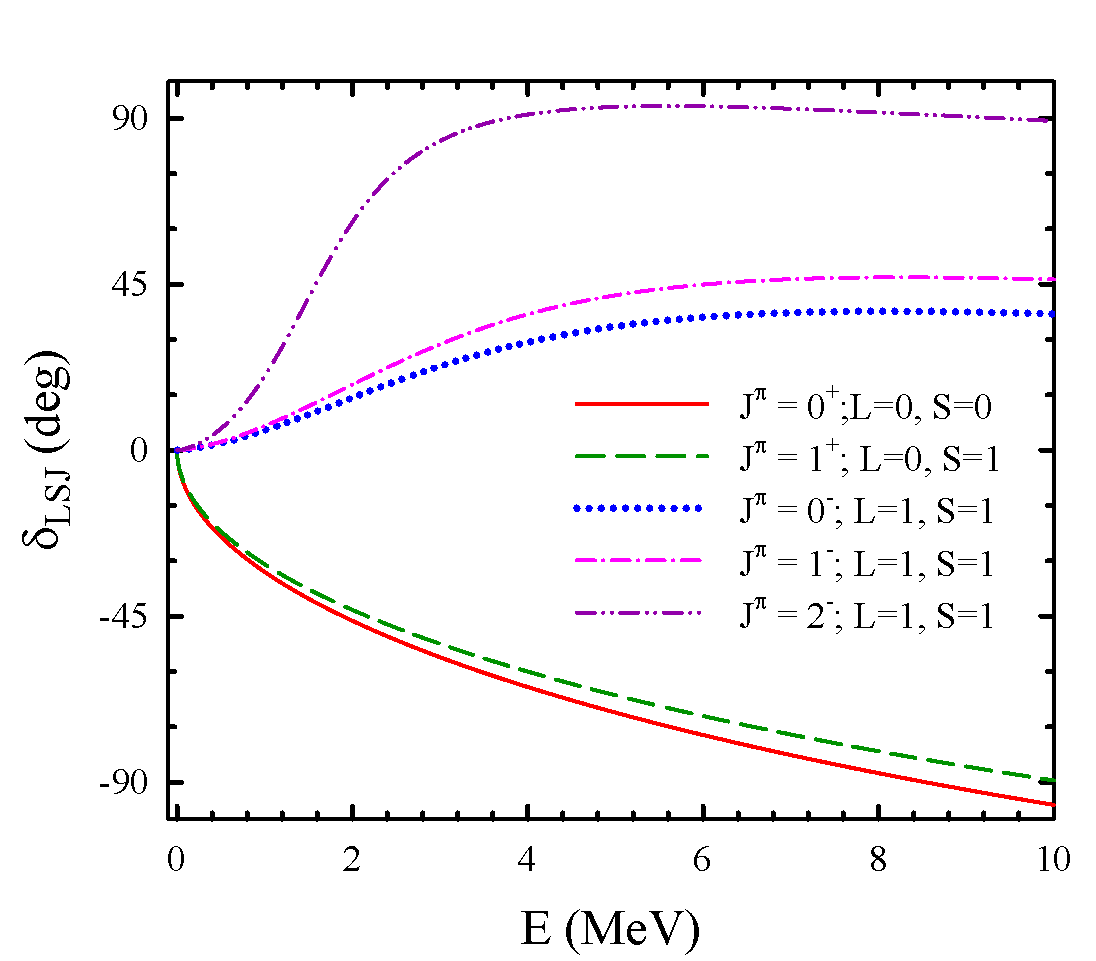}%
\caption{Phase shifts of the elastic neutron scattering on $^{3}$H.}%
\label{Fig:PhasesN3H}%
\end{center}
\end{figure}

The phase shifts of the elastic scattering of neutron on $^{3}$H are shown in Fig.
\ref{Fig:PhasesN3H}. Phase shifts in the states $J^{\pi}$=2$^{-}$, 1$^{-}$ and
0$^{-}$ ($L$=1, $S$=1) exhibit resonance behavior. The detected resonance states
are quite wide, which is in agreement with the experimental observations
\cite{1992NuPhA.541....1T}. One also notices similarity in the behavior of $\Lambda$+$^{3}$H and $n$+$^{3}$H scattering at the states $J^{\pi}$=0$^+$, $L=S=0$ and $J^{\pi}$=1$^+$, $L=0$, $S=0$. These phase shifts are decreasing very fast with an increase in energy.

\begin{figure}
[ptb]
\begin{center}
\includegraphics[
height=15.752cm,
width=13.5026cm
]%
{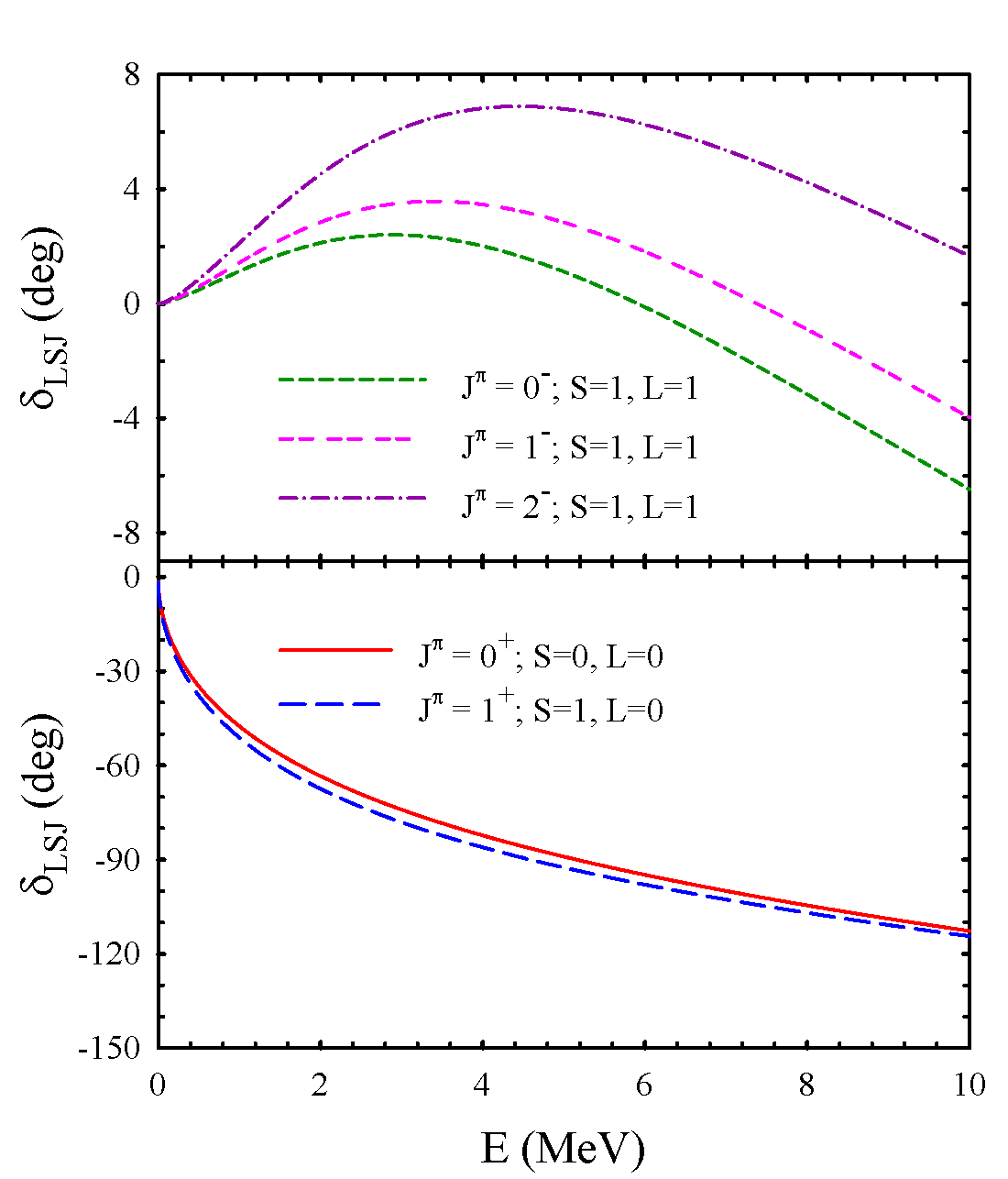}%
\caption{Phase shifts of the elastic $\Lambda$+$^{3}$He scattering.}%
\label{Fig:PhasesL3He}%
\end{center}
\end{figure}

\begin{figure}[ptb]
\begin{center}
\includegraphics[
height=11.7893cm,
width=13.4917cm
]{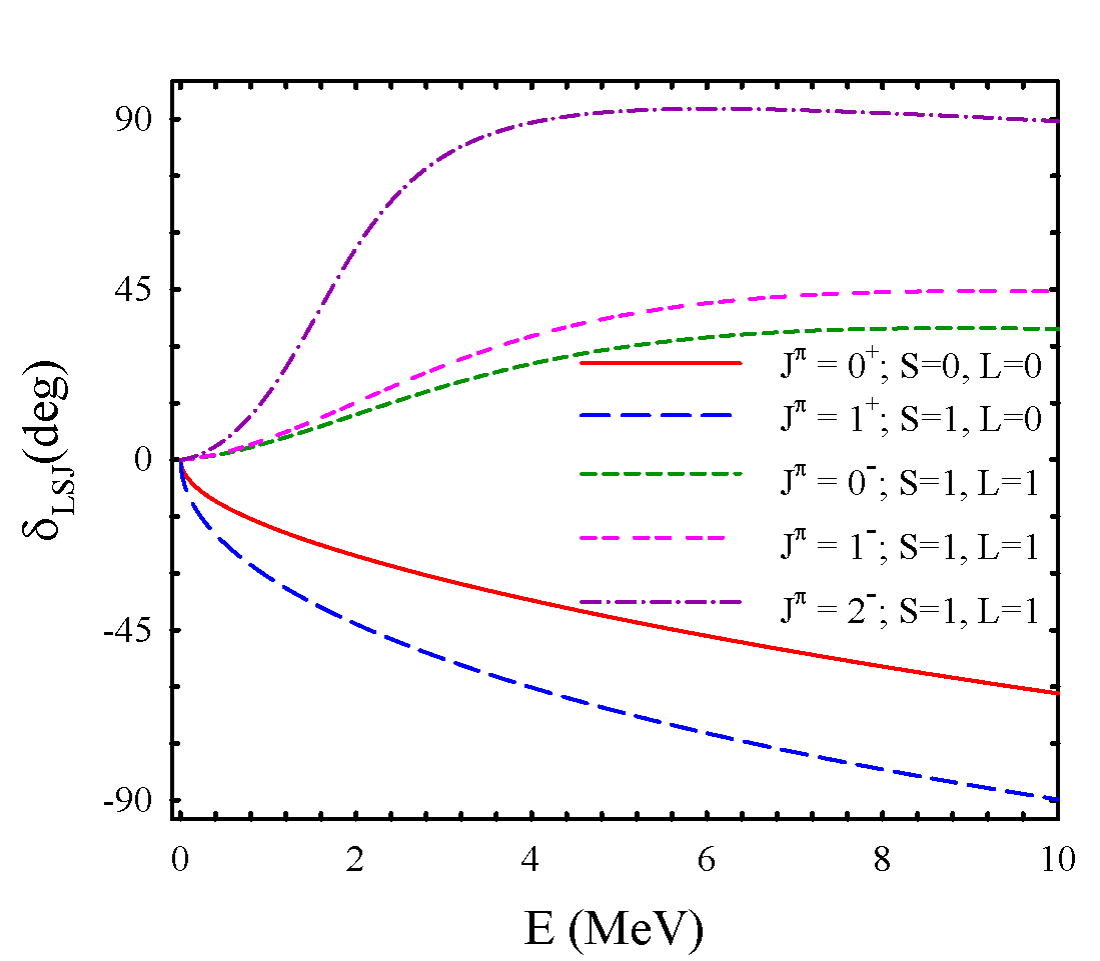}%
\caption{Phase shifts of the elastic neutron scattering on $^{3}$He.}%
\label{Fig:PhasesN3He}%
\end{center}
\end{figure}

In Figs. \ref{Fig:PhasesL3He} and \ref{Fig:PhasesN3He} we show behavior of
phase shifts of the elastic $\Lambda$+$^{3}$He and $n$+$^{3}$He scattering.
Similarly to the cases of $\Lambda$+$^{3}$H and $n$+$^{3}$H, the 0$^{+}$ and
1$^{+}$ phase shifts are rapidly decreasing.  For $\Lambda$+$^{3}$He system
such behavior is stipulated by the existence of the corresponding 0$^{+}$ and
1$^{+}$ bound states, while there is only bound \ state in the 0$^{+}$ state
of $n$+$^{3}$He system. The slow growing of phase shifts of the elastic
$\Lambda$+$^{3}$He scattering with the total orbital momentum $L$=1 and three
angular momenta 0$^{-}$, 1$^{-}$ and 2$^{-}$ indicates that there are no
resonance states of negative parity in $_{\Lambda}^{4}$He.%

%

\begin{figure}[hptb]
\begin{center}
\includegraphics[width=\textwidth]{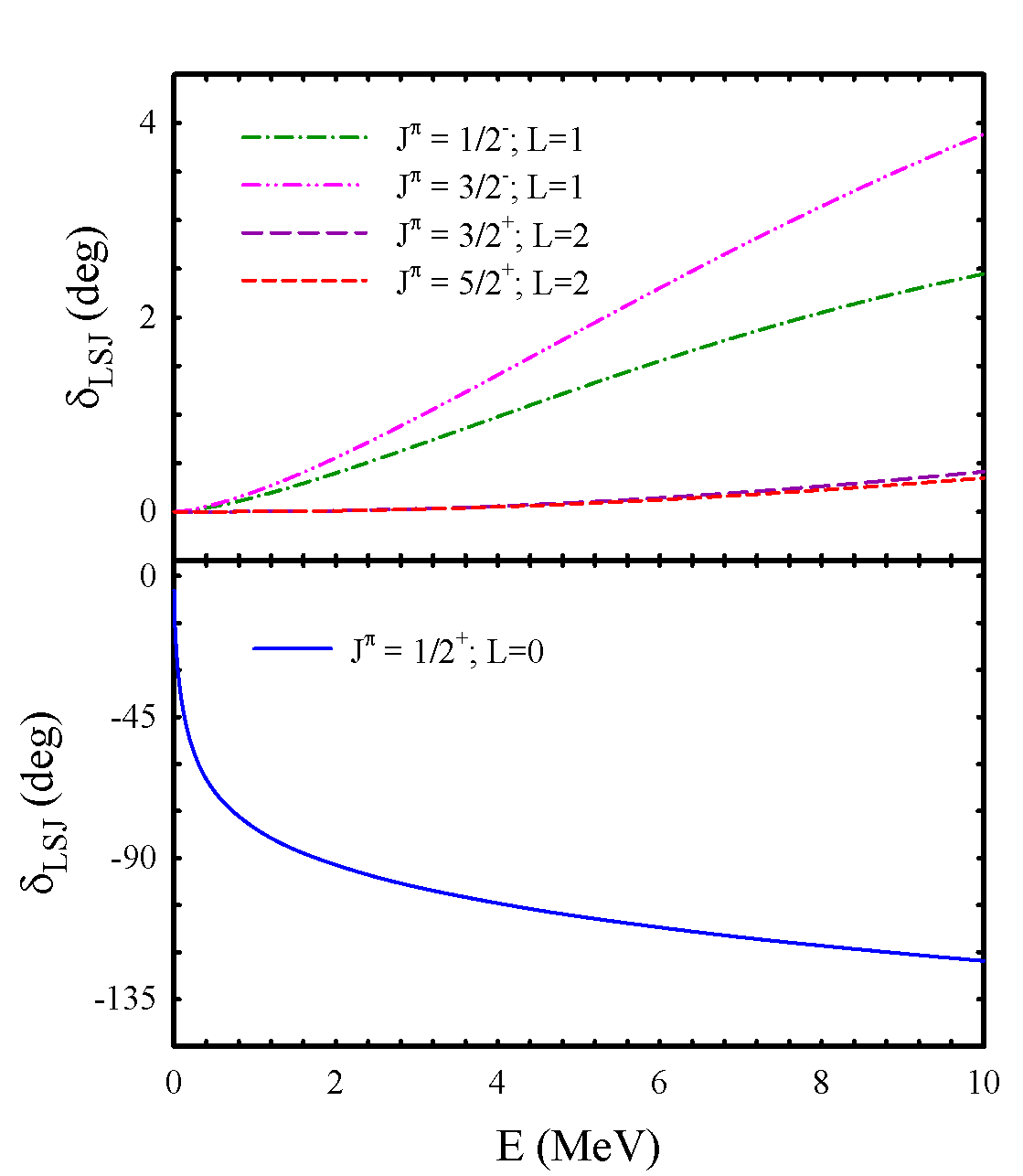}%
\caption{Phase shifts of the $\Lambda$ elastic scattering on deuteron.}%
\label{Fig:PhasesL2H}%
\end{center}
\end{figure}

\begin{figure}[hptb]
\begin{center}
\includegraphics[width=\textwidth]{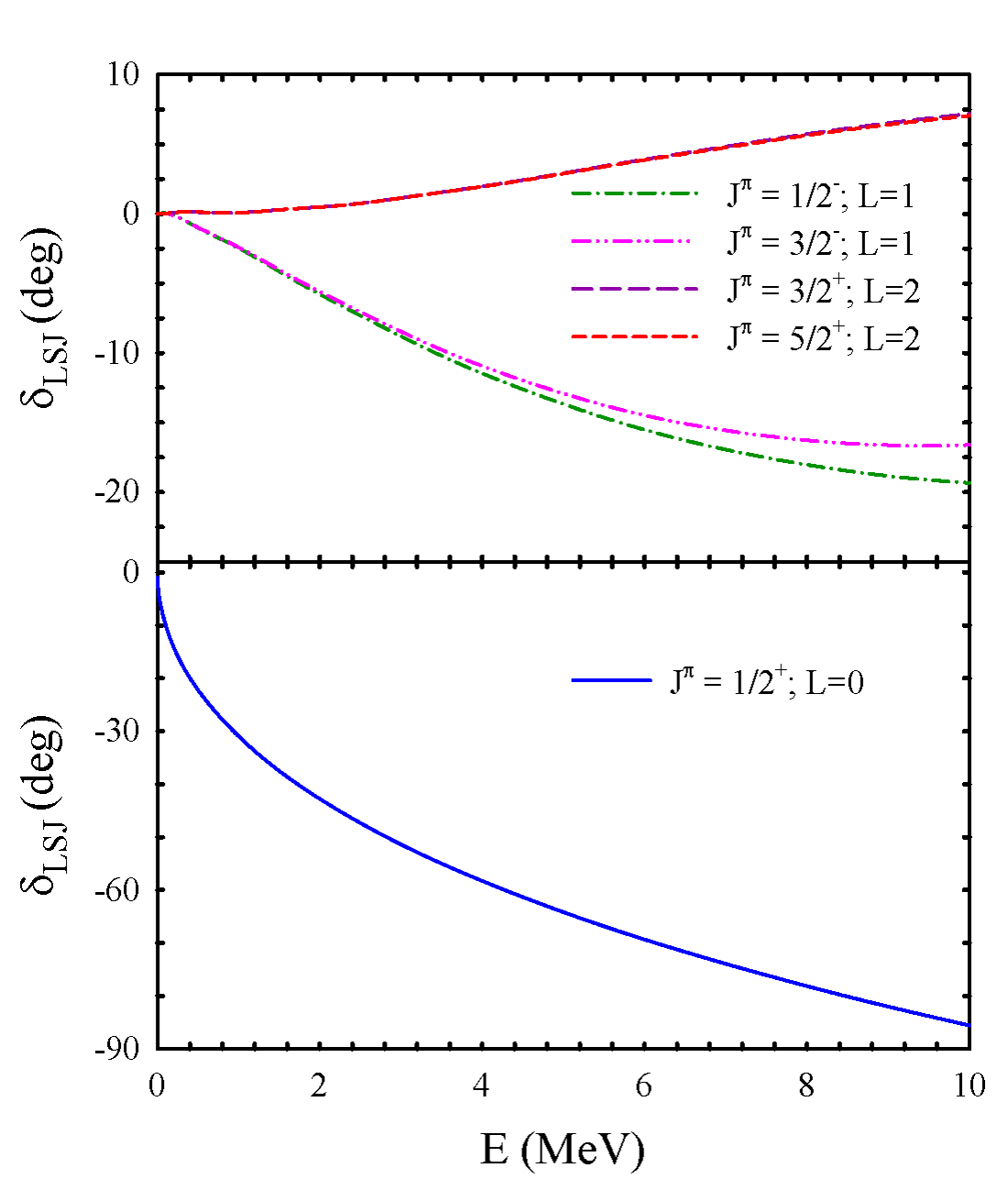}%
\caption{Phase shifts of the elastic neutron scattering on deuteron.}%
\label{Fig:Phasesn2H}%
\end{center}
\end{figure}

Now we consider the scattering of the lambda hyperon and the neutron on the deuteron. The phase shifts of elastic $\Lambda + d$ and $n  + d$ are displayed in Figs. \ref{Fig:PhasesL2H} and \ref{Fig:Phasesn2H}, respectively. As in previous cases, the s-wave scattering phase shifts decrease with increasing energy. Recall that both $_{\Lambda}^3$H and $^3$H have a bound state, and the existence of a bound state reflects the behavior of the corresponding phase shifts. We can also see that the centrifugal barrier is small in the hypernucleus and nucleus, and it does not create a resonance state. Figs. \ref{Fig:PhasesL2H} and \ref{Fig:Phasesn2H} also demonstrate that the interaction of lambda hyperon with deuteron in states with the total orbital momenta $L$=1  and $L$=2 is stronger than the interaction of neutron with deuteron in these states, as the phase shifts of the $\Lambda + d$  scattering are substantially larger than the phase shifts of the $n  + d$ scattering.

\begin{figure}[hptb]
\begin{center}
\includegraphics[width=\textwidth]{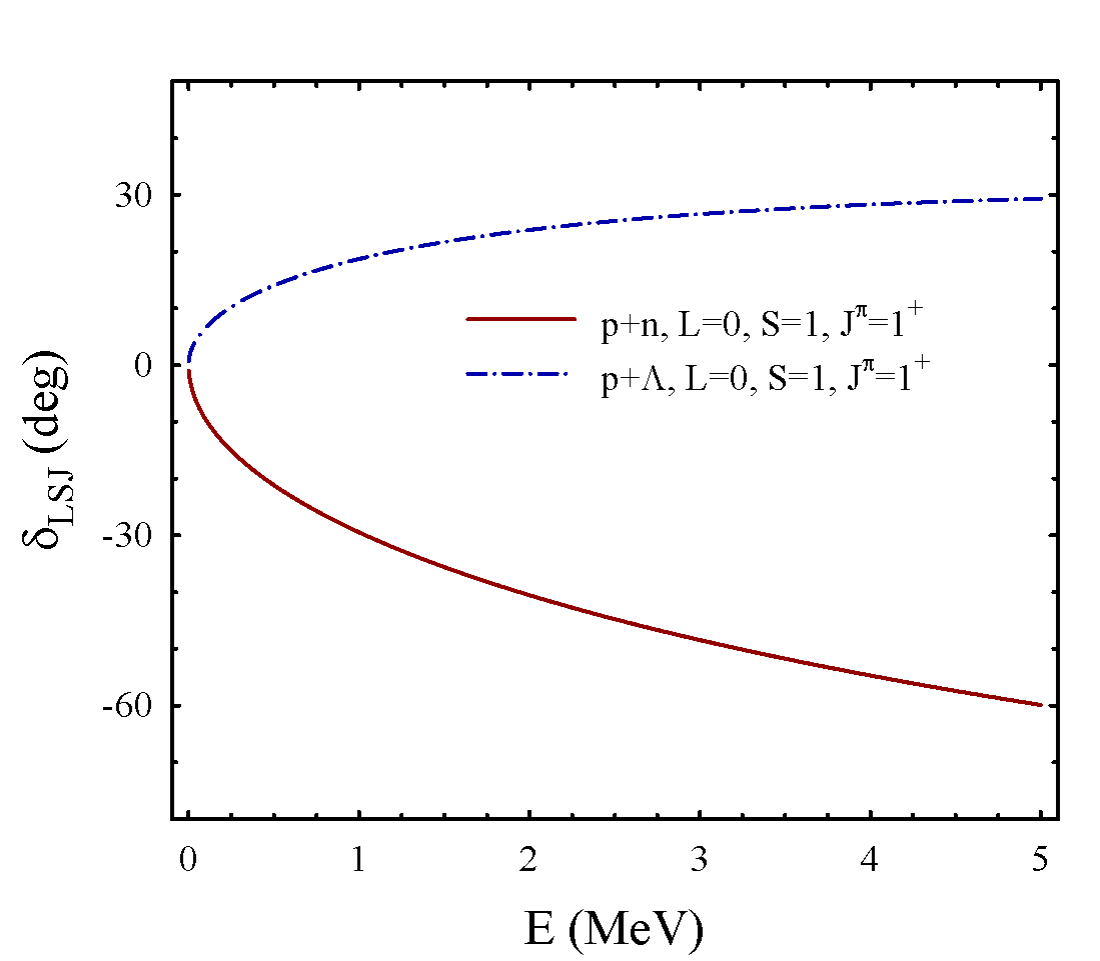}%
\caption{Phase shits of the elastic $p+n$ and $p+\Lambda$ scattering in the
state $L$=0, $S$=1 and $J^{\pi}$=1$^{+}$.}%
\label{Fig:PhasesPN_PL}%
\end{center}
\end{figure}

In Fig. \ref{Fig:PhasesPN_PL} we display the phase shifts of the elastic
scattering of lambda hyperon and neutron on proton in the state $J^{\pi}%
$=1$^{+}$ (i.e. with the total orbital momentum $L$=0 and total spin $S$=1).
There is the strongest interaction between the proton and neutron in this state,
as it creates the bound 1$^{+}$ state. This bound state is reflected in
behavior of the phase shift which rapidly decreases with increasing of the
energy. The phase shifts of the elastic $p+\Lambda$ scattering also shows that
the  $p+\Lambda$ interaction in the 1$^{+}$ state has an attractive character,
however, this attraction is too week to create a bound state or resonance state.%

\subsection{Scattering length}

Let us consider low-energy phase shifts in more detail.  It is well known
\cite{kn:Newton} (Sec. 11, p. 317), \cite{kn:kukulin89} (Sec. 6.1, p. 281)  that in this energy region
\begin{equation}
k^{2L+1}\cot\delta_{L}\approx-\frac{1}{a_{L}}+\frac{1}{2}r_{L}k^{2}-\frac
{1}{4}P_{L}k^{4} \label{eq:S041},%
\end{equation}
where $a_{L}$ is the scattering length, $r_{L}$ is the effective range,
$P_{L}$ is the form parameter. We determined these low-energy parameters for
lambda hyperon scattering with zero orbital momentum on  the s-shell nuclei and
show them in Table \ref{Tab:ScattLength}. In Table \ref{Tab:ScattLength} we
also show the scattering length, effective range and form parameter for
neutron scattering on the s-shell nuclei.%

\begin{table}[ht] \centering
\begin{ruledtabular}  
\caption{Parameters of low-energy scattering (scattering length $a_{L}$, the effective range $r_{L}$,   and the shape parameter $P_{L}$) of lambda hyperon and neutron on lightest nuclei. Energy is given in MeV, the scattering length and effective range are in fm, and the form parameter is in fm$^{3}$.\label{Tab:ScattLength}}%
\begin{tabular}
[c]{ccccccccc}
Channel & $E_{BS}$ & $a_{0}$ & $r_{0}$ & $P_{0}$ & Channel & $a_{0}$ & $r_{0}$
		& $P_{0}$\\\hline
		$\Lambda$+$d$ & -0.164 & 14.26 & 2.27 & -0.04 & $n$+$d$ & 2.92 & 13.26 &
		0.24\\\hline
		$\Lambda$+$t$, $S$=0 & -2.165 & 4.68 & 1.95 & -0.02 & $n$+$t$, $S$=0 & 3.01 &
		4.22 & 1.22\\\hline
		$\Lambda$+$t$, $S$=1 & -1.623 & 5.23 & 2.06 & -0.03 & $n$+$t$, $S$=1 & 2.81 &
		5.07 & 1.03\\\hline
$\Lambda$+$^{3}$He, $S$=0 & -2.346 & 4.54 & 1.93 & -0.02 & $n$+$^{3}$He,
$S$=0 & 1.48 & 25.28 & 0.06 \\\hline
$\Lambda$+$^{3}$He, $S$=1 & -1.787 & 5.05 & 2.03 & -0.03 & $n$+$^{3}$He,
$S$=1 & 2.82 & 5.24 & 0.97 \\\hline
		$\Lambda$+$^{4}$He & -3.104 & 4.10 & 1.92 & -0.03 & $n$+$^{4}$He & 2.46 &
		4.41 & 1.29\\\hline
\end{tabular}
\end{ruledtabular}  
\end{table}%
    
\begin{figure}[hptb]
\begin{center}
\includegraphics[width=\textwidth]{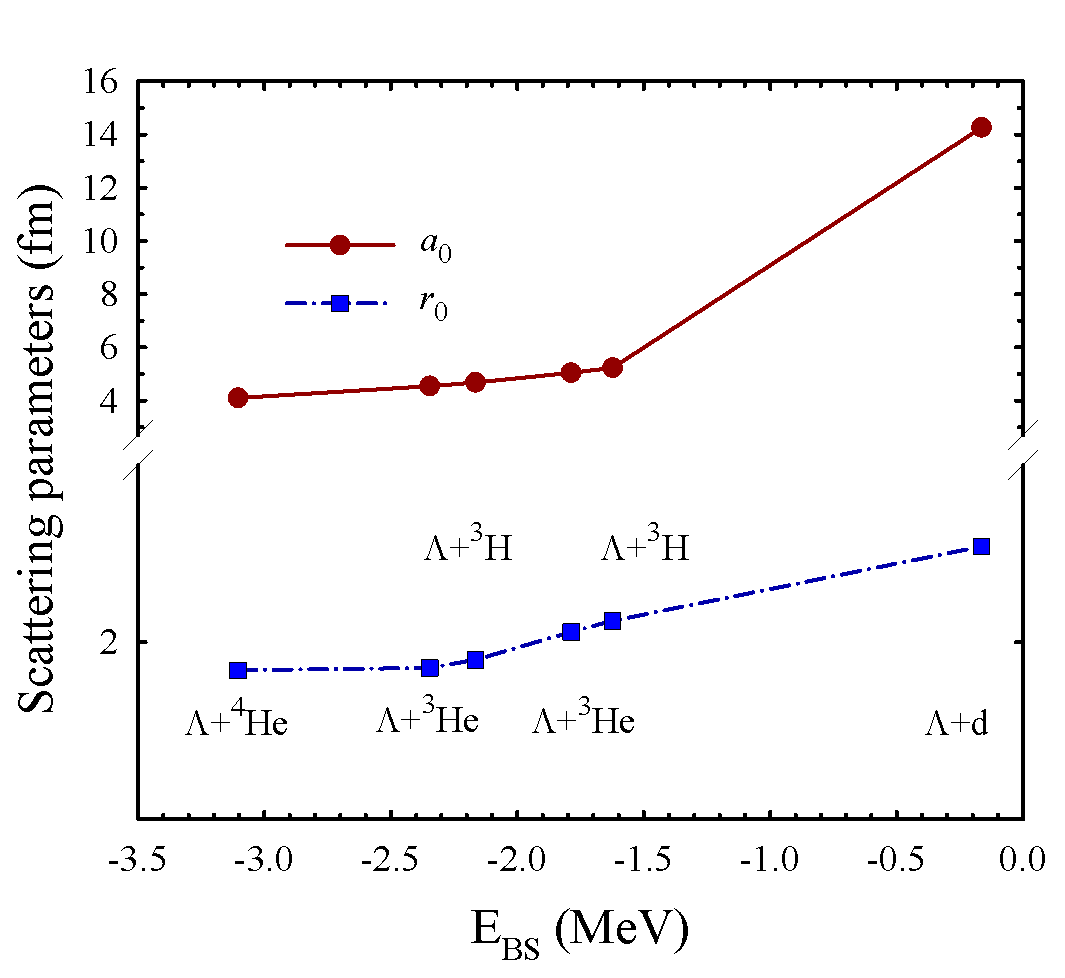}%
\caption{Scattering length  $a_{0}$ and effective range $r_{0}$ as a function of the bound state energy of light hypernuclei.}%
\label{Fig:ScattLengthvsEBS}%
\end{center}
\end{figure}

In Fig. \ref{Fig:ScattLengthvsEBS} we demonstrate the correlation between the
scattering length $a_{0}$ and the energy $E_{BS}$ of the bound state of a
hypernucleus. As we can see, the smaller the energy $\left\vert
E_{BS}\right\vert $ of the bound state, the larger is the scattering length.
Indeed, the largest scattering 13.59 fm is obtained for $_{\Lambda}^{3}$H,
where the bound-state energy is $E_{BS}=$-0.177 MeV, and the smallest
scattering length 3.95 fm is found in $_{\Lambda}^{5}$He, the hypernucleus
with the deepest bound state $E_{BS}$=-3.20 MeV. In Fig.
\ref{Fig:ScattLengthvsEBS} we also show how the effective range $r_{0}$
depends on the energy of the bound state. Similarly to the scattering length, the
effective range  slightly decreases with the increase of the bound-state energy $\left\vert E_{BS}\right\vert $. 

In Table \ref{Tab:ScattLength} we also display the scattering length, the
effective range and the form parameter for the elastic scattering of neutrons
on the s-shell nuclei. However, we do not display bound-state energies of ordinary nuclei as the only
two nuclei ($^{3}$H=$d+n$ and $^{4}$He=$^3$He+$n$) are bound. One can see that in all cases, displayed
in Table \ref{Tab:ScattLength}, the neutron scattering length is smaller than the
lambda hyperon scattering length. Besides, the scattering parameters for the
$\Lambda$+$t$ and $\Lambda$+$^{4}$He elastic scattering are comparable to the
parameters of the $n$+$t$ and $n$+$^{4}$He.%

\subsection{Comparing with other model}

In Ref. \cite{2002PhRvL..89n2504N}, the hypernuclei $_{\Lambda}^{3}$H,
$_{\Lambda}^{4}$H and $_{\Lambda}^{5}$He have been studied in ab initio
calculations with the YN potentials, which involve spin-orbit and tensor
components. Four sets of the hyperon-nucleon interactions were used in these calculations. Such
potentials, taking into account the $\Lambda$N - $\Sigma$N coupling, were suggested in
Ref. \cite{2000PhRvL..84.3539A} and  were
denoted as SC97d(S), SC97e(S), SC97f(S) and SC89(S). In Table \ref{Tab:NemuravsOurResults} we
show results obtained in Ref. \cite{2002PhRvL..89n2504N} for bound states of
$_{\Lambda}^{3}$H, $_{\Lambda}^{4}$H and $_{\Lambda}^{5}$He with four used
potentials. We also present the same results of our calculations, denoted as YNG-NF. Similarly to
Ref. \cite{2002PhRvL..89n2504N}, we also calculated the singlet ($a_{s}$) and
triplet ($a_{t}$) $\Lambda$-N scattering lengths. Recall that we used only one
$\Lambda$N potential  but with different value of the parameter $k_{F}$, thus
we display three sets of $a_{s}$ and $a_{t}$ scattering lengths. One can see
that the energy of bound states ($E_{BS}$) and energy of exited state
($E_{ES}$) in $_{\Lambda}^{4}$H strongly depends on the shape of $\Lambda$N
potentials, employed in Ref. \cite{2002PhRvL..89n2504N}. These potentials
generates also fairly different values of the singlet and triplet scattering
lengths. In our calculations, the singlet scattering length is always larger
than the triplet scattering length; the same is observed in
\cite{2002PhRvL..89n2504N} for all but one case.%

\begin{table}[ht] \centering
\begin{ruledtabular}  
\caption{Spectrum of $_{\Lambda }^{3}$H, $_{\Lambda }^{4}$H and $_{\Lambda}^{5}$He hypernuclei obtained in two different models.\label{Tab:NemuravsOurResults}}%
\begin{tabular}
[c]{ccccccc}
Potential & $a_{s}$ & $a_{t}$ & $E_{BS}$($_{\Lambda}^{3}$H) & $E_{BS}$%
		($_{\Lambda}^{4}$H) & $E_{ES}$($_{\Lambda}^{4}$H) & $E_{BS}$($_{\Lambda}^{5}%
		$He)\\\hline
		SC97d(S), \cite{2002PhRvL..89n2504N}  & -1.92 & -1.96 & -0.01 & -1.67 & -1.20 &
		-3.17\\
		SC97e(S), \cite{2002PhRvL..89n2504N} & -2.37 & -1.83 & -0.10 & -2.06 & -0.92 &
		-2.75\\
		SC97f(S), \cite{2002PhRvL..89n2504N} & -2.82 & -1.72 & -0.18 & -2.16 & -0.63 &
		-2.10\\
		SC89(S), \cite{2002PhRvL..89n2504N} & -3.39 & -1.38 & -0.37 & -2.55 & - &
		-0.35\\\hline
		& -3.09 & -2.39 & -0.16 &  &  & \\
		YNG-NF & -2.95 & -2.28 &  & -2.17 & -1.62 & \\
		& -2.47 & -1.89 &  &  &  & -3.10\\\hline
\end{tabular}
\end{ruledtabular}  
\end{table}%

One may conclude that there are consistencies between the results of the two-cluster
model and ab initio calculations for the lightest hypernuclei.

\section{Conclusions \label{Sec:Conclusions}}

We studied the peculiarities of the interaction of the lambda hyperon with the s-shell nuclei and compared them with the interaction of the neutron with the same nuclei. These studies are carried out with the algebraic version of the resonating group method, which involves a full set of harmonic oscillator functions to expand
wave functions of the relative motion of interacting clusters. It was demonstrated
that the present model fairly well reproduces the spectrum of bound states in hypernuclei of interest.  

It was also demonstrated that there are no resonance states in the two-cluster
continuum formed by the interaction of the lambda hyperon with the s-shell nuclei. It
means that nucleon-lambda interaction is weak and consequently, s-nucleus-lambda interaction
is also weak to create a favorable situation for forming a resonance state.

We concluded that the interaction of the lambda hyperon with the s-shell nuclei is weaker than the interaction of neutron with the same nuclei, however, the lack of the Pauli principle in the system, comprised of lambda hyperon and nucleons, leads to formation of bound states in $_{\Lambda}^{4}$H, $_{\Lambda}^{5}$He, where their counterparts have no bound states.

We compared our results, obtained from a two-cluster model, with the results of other models and demonstrated the consistency of our model with the three- and four-cluster models.

The results obtained in this paper are used in Ref. \cite{2025arXiv250813702K}  
to study spectrum of bound and resonance states of $_{\Lambda}^{7}$Li
hypernucleus, considered as a three-cluster system $^{4}$He+$d$+$\Lambda$. It
is also planned to study hypernuclei $_{\Lambda}^{8}$Li and $_{\Lambda}^{8}%
$Be, where the interaction of the lambda hyperon with s-shell nuclei plays an
important role.

\begin{acknowledgments}

This work was partially supported by the Science Committee of the
	Ministry of Education and Science of the Republic of Kazakhstan (Grant No.
	AP22683187, the project title "Structure of the light nuclei and hypernuclei
	in multi-channel and multi-cluster models") and received partial support from
	the Program of Fundamental Research of the Physics and Astronomy Department of
	the National Academy of Sciences of Ukraine (Project No. 0122U000889). V.V.S.
	extends his gratitude to the Simons Foundation for their financial support
	(Award ID: SFI-PD-Ukraine-00014580).

\end{acknowledgments}

\bibliography{ASL}

@misc{ChartHyperN2021,
  title        = {Chart of hypernuclides -- Hypernuclear structure and decay data},
  author       = {{Eckert}, P. and {Achenbach}, P. and {et al.}},
  year         = 2021,
  note         = {https://hypernuclei.kph.uni-mainz.de}
}

@ARTICLE{2025ChPhL..42j0101C,
       author = {{Chen}, Jin-Hui and {Geng}, Li-Sheng and {Hiyama}, Emiko and {Liu}, Zhi-Wei and {Pochodzalla}, Josef},
        title = "{Perspectives for Hyperon and Hypernuclei Physics}",
      journal = {Chin. Phys. Lett.},
         year = 2025,
        month = sep,
       volume = {42},
       number = {10},
          eid = {100101},
        pages = {100101},
          doi = {10.1088/0256-307X/42/10/100101},
archivePrefix = {arXiv},
       eprint = {2506.00864},
 primaryClass = {nucl-th},
       adsurl = {https://ui.adsabs.harvard.edu/abs/2025ChPhL..42j0101C}
}

@ARTICLE{2025PhRvL.134g2502L,
       author = {{Le}, Hoai and {Haidenbauer}, Johann and {Mei{\ss}ner}, Ulf-G. and {Nogga}, Andreas},
        title = "{Light $\Lambda$ Hypernuclei Studied with Chiral Hyperon-Nucleon and Hyperon-Nucleon-Nucleon Forces}",
      journal = {Phys. Rev. Lett.},
         year = 2025,
        month = feb,
       volume = {134},
       number = {7},
          eid = {072502},
        pages = {072502},
          doi = {10.1103/PhysRevLett.134.072502},
archivePrefix = {arXiv},
       eprint = {2409.18577},
 primaryClass = {nucl-th},
       adsurl = {https://ui.adsabs.harvard.edu/abs/2025PhRvL.134g2502L}
}

@ARTICLE{2010NuPhA.848....1P,
   author = {{Purcell}, J.~E. and {Kelley}, J.~H. and {Kwan}, E. and {Sheu}, C.~G. and
    {Weller}, H.~R.},
    title = "{Energy levels of light nuclei A=3}",
  journal = {Nucl. Phys. A},
     year = 2010,
    month = dec,
   volume = 848,
    pages = {1-74},
      doi = {10.1016/j.nuclphysa.2010.08.012},
   adsurl = {http://adsabs.harvard.edu/abs/2010NuPhA.848....1P}
}

@ARTICLE{2002NuPhA.708....3T,
   author = {{Tilley}, D.~R. and {Cheves}, C.~M. and {Godwin}, J.~L. and
    {Hale}, G.~M. and {Hofmann}, H.~M. and {Kelley}, J.~H. and {Sheu}, C.~G. and
    {Weller}, H.~R.},
    title = "{Energy levels of light nuclei  \makebox{$A$}=5, 6, 7}",
  journal = {Nucl. Phys. A},
     year = 2002,
    month = sep,
   volume = 708,
    pages = {3-163},
      doi = {10.1016/S0375-9474(02)00597-3},
   adsurl = {http://adsabs.harvard.edu/cgi-bin/nph-bib_query?bibcode=2002NuPhA.708....3T&db_key=PHY}
}

@ARTICLE{1992NuPhA.541....1T,
   author = {{Tilley}, D.~R. and {Weller}, H.~R. and {Hale}, G.~M.},
    title = "{Energy levels of light nuclei \makebox{$A$} = 4}",
  journal = {Nucl. Phys. A},
     year = 1992,
    month = may,
   volume = 541,
    pages = {1-104},
      doi = {10.1016/0375-9474(92)90635-W},
   adsurl = {http://adsabs.harvard.edu/cgi-bin/nph-bib_query?bibcode=1992NuPhA.541....1T&db_key=PHY}
}

@ARTICLE{2002PhRvL..89n2504N,
       author = {{Nemura}, H. and {Akaishi}, Y. and {Suzuki}, Y.},
        title = "{Ab initio Approach to s-Shell Hypernuclei $^{3}$$_{{\ensuremath{\Lambda}}}$H, $^{4}$$_{{\ensuremath{\Lambda}}}$H, $^{4}$$_{{\ensuremath{\Lambda}}}$He, and $^{5}$$_{{\ensuremath{\Lambda}}}$He with a {\ensuremath{\Lambda}}N-{\ensuremath{\Sigma}}N Interaction}",
      journal = {Phys. Rev. Lett.},
         year = 2002,
        month = sep,
       volume = {89},
       number = {14},
          eid = {142504},
        pages = {142504},
          doi = {10.1103/PhysRevLett.89.142504},
archivePrefix = {arXiv},
       eprint = {nucl-th/0203013},
 primaryClass = {nucl-th},
       adsurl = {https://ui.adsabs.harvard.edu/abs/2002PhRvL..89n2504N}
}

@ARTICLE{kn:Fil_Okhr,
      AUTHOR=    "G.~ F.~ Filippov and I.~ P.~ Okhrimenko",
      TITLE=     "Use of an oscillator basis for solving continuum problems",
      JOURNAL=   "Sov. J. Nucl. Phys.",
      YEAR=      "1981",
      VOLUME=    "{\bf 32}",
      PAGES=     "480-484"
      }

@ARTICLE{kn:Fil81,
      AUTHOR=    "G.~ F.~ Filippov",
      TITLE=     "On taking into account correct asymptotic behavior in oscillator-basis expansions",
      JOURNAL=   "Sov. J. Nucl. Phys.",
      YEAR=      "1981",
      VOLUME=    "{\bf 33}",
      PAGES=     "488-489"
      }

@ARTICLE{2021NuPhA101622325N,
       author = {{Nesterov}, A.~V. and {Lashko}, Yu. A. and {Vasilevsky}, V.~S.},
        title = "{Structure of the ground and excited states in $_{\Lambda}^{9}$Be nucleus}",
      journal = {Nucl. Phys. A},
         year = 2021,
        month = dec,
       volume = {1016},
          eid = {122325},
        pages = {122325},
          doi = {10.1016/j.nuclphysa.2021.122325},
       adsurl = {https://ui.adsabs.harvard.edu/abs/2021NuPhA101622325N}
}

@ARTICLE{1937PhRv...52.1083W,
   author = {{Wheeler}, J.~A.},
    title = "{Molecular Viewpoints in Nuclear Structure}",
  journal = {Phys. Rev.},
     year = 1937,
    month = dec,
   volume = 52,
    pages = {1083-1106},
      doi = {10.1103/PhysRev.52.1083},
   adsurl = {http://adsabs.harvard.edu/abs/1937PhRv...52.1083W}
}

@ARTICLE{1937PhRv...52.1107W,
   author = {{Wheeler}, J.~A.},
    title = "{On the Mathematical Description of Light Nuclei by the Method of Resonating Group Structure}",
  journal = {Phys. Rev.},
     year = 1937,
    month = dec,
   volume = 52,
    pages = {1107-1122},
      doi = {10.1103/PhysRev.52.1107},
   adsurl = {http://adsabs.harvard.edu/abs/1937PhRv...52.1107W}
}

@ARTICLE{potMHN1,
      AUTHOR=   "A. Hasegawa and S. Nagata",
      TITLE=    "Ground State of  \makebox{$^6$Li}",
      JOURNAL=  "Prog. Theor. Phys.",
      YEAR=     "1971",
      VOLUME=   "{45}",
      PAGES=    "1786-1807",
      doi = {10.1143/PTP.45.1786},
   adsurl = {http://adsabs.harvard.edu/abs/1971PThPh..45.1786H}
      }

@ARTICLE{potMHN2,
      AUTHOR=   "F. Tanabe and A. Tohsaki and R. Tamagaki",
      TITLE=    "$\alpha \alpha$ scattering at intermediate energies",
      JOURNAL=  "Prog. Theor. Phys.",
      YEAR=     "1975",
      VOLUME=   "{53}",
      PAGES=    "677-691",
        doi = {10.1143/PTP.53.677},
       adsurl = {https://ui.adsabs.harvard.edu/abs/1975PThPh..53..677T}
}

@ARTICLE{1994PThPS.117..361Y,
       author = {{Yamamoto}, Y. and {Motoba}, T. and {Himeno}, H. and {Ikeda}, K. and {Nagata}, S.},
        title = "{Hyperon-Nucleon and Hyperon-Hyperon Interactions in Nuclei}",
      journal = {Prog. Theor. Phys. Suppl.},
         year = 1994,
        month = jan,
       volume = {117},
        pages = {361-389},
          doi = {10.1143/PTPS.117.361},
       adsurl = {https://ui.adsabs.harvard.edu/abs/1994PThPS.117..361Y}
}

@ARTICLE{1979PhR....55..183S,
       author = {{Satchler}, G.~R. and {Love}, W.~G.},
        title = "{Folding model potentials from realistic interactions for heavy-ion scattering}",
      journal = {Phys. Rep.},
         year = 1979,
        month = oct,
       volume = {55},
       number = {3},
        pages = {183-254},
          doi = {10.1016/0370-1573(79)90081-4},
       adsurl = {https://ui.adsabs.harvard.edu/abs/1979PhR....55..183S}
}

@ARTICLE{Nesterov:2021gcp,
       author = {{Nesterov}, A. V. and {Solokha-Klymchak}, M.},
        title = "{Properties of $^4_\Lambda$H Hypernucleus in Three-Cluster Microscopic Models}",
     journal = {Ukr. J. Phys},
        volume = {66}, 
       number = {10}, 
        pages = {846-856},
        year = "2021",
         doi = {10.15407/ujpe66.10.846}
}

@BOOK{kn:Newton,
      AUTHOR=    "R.~G.~ Newton",
      TITLE=     "Scattering Theory of Waves and Particles",
      PUBLISHER=   "McGraw-Hill",
      YEAR=      "1966",
      ADDRESS=    "New-York",
      PAGES=     ""
      }

@article{DUISENBAY2019Bul,
title = "{Form factors and density distributions of protons and neutrons in $^{7}$Li and  $^{7}$Be}",
journal = "News Nat. Acad. Scien. Rep. Kazakhstan",
volume = "3",
number = "325",
pages = "71-76",
year = "2019",
issn = "1991-346X",
doi = "https://doi.org/10.32014/2019.2518-1726.26",
author = "{A. D. Duisenbay and N. Zh. Takibayev and V. S. Vasilevsky and V. O. Kurmangaliyeva and E. M. Akzhigitova}"
}

@ARTICLE{2015NuPhA.941..121L,
   author = {{Lashko}, Y.~A. and {Filippov}, G.~F. and {Vasilevsky}, V.~S.},
    title = "{Dynamics of two-cluster systems in phase space}",
  journal = {Nucl. Phys. A},
archivePrefix = "arXiv",
   eprint = {1503.06005},
 primaryClass = "nucl-th",
     year = 2015,
    month = sep,
   volume = 941,
    pages = {121-144},
      doi = {10.1016/j.nuclphysa.2015.06.006},
   adsurl = {http://adsabs.harvard.edu/abs/2015NuPhA.941..121L}
}

@ARTICLE{kn:cohstate1E,
      AUTHOR=    "G.~F. Filippov and V.~S. Vasilevsky and L.~L. Chopovsky",
      TITLE=     "Generalized coherent states in nuclear-physics problems",
      JOURNAL=   "Sov. J. Part. Nucl.",
      YEAR=      "1984",
      VOLUME=    "{\bf 15}",
      PAGES=     "600-619"
      }

@ARTICLE{kn:cohstate2E,
      AUTHOR=    "G.~F. Filippov and V.~S. Vasilevsky and L.~L. Chopovsky",
      TITLE=     "Solution of problems in the microscopic theory of the nucleus using the
                  technique of generalized coherent states",
      JOURNAL=   "Sov. J. Part. Nucl.",
      YEAR=      "1985",
      VOLUME=    "{\bf 16}",
      PAGES=     "153-177"
      }

@ARTICLE{2023UkrJPh..68..3K,
   author = {{Kalzhigitov}, N. and {Kurmangaliyeva}, V.~O. and  {Takibayev}, N.~Zh. and  {Vasilevsky}, V.~S.},
    title = "{Resonance structure of $^8$Be within the two-cluster resonating group method}",
  journal = {Ukr. J. Phys.},
     year = 2023,
   volume = {68},
   number =  {1},
    pages = {3-18},
    doi = {10.15407/ujpe68.1.3}
}

@ARTICLE{2000PhRvL..84.3539A,
       author = {{Akaishi}, Y. and {Harada}, T. and {Shinmura}, S. and {Myint}, Khin Swe},
        title = "{Coherent {\ensuremath{\Lambda}}-{\ensuremath{\Sigma}} Coupling in s-Shell Hypernuclei}",
      journal = {Phys. Rev. Lett.},
         year = 2000,
        month = apr,
       volume = {84},
       number = {16},
        pages = {3539-3541},
          doi = {10.1103/PhysRevLett.84.3539},
       adsurl = {https://ui.adsabs.harvard.edu/abs/2000PhRvL..84.3539A}
}

@ARTICLE{2001PhRvC..65a1301H,
       author = {{Hiyama}, E. and {Kamimura}, M. and {Motoba}, T. and {Yamada}, T. and {Yamamoto}, Y.},
        title = "{{\ensuremath{\Lambda}}-{\ensuremath{\Sigma}} conversion in $^{4}$$_{{\ensuremath{\Lambda}}}$He and $^{4}$$_{{\ensuremath{\Lambda}}}$H based on a four-body calculation}",
      journal = {Phys. Rev. C},
         year = 2001,
        month = dec,
       volume = {65},
       number = {1},
          eid = {011301},
        pages = {011301},
          doi = {10.1103/PhysRevC.65.011301},
archivePrefix = {arXiv},
       eprint = {nucl-th/0106070},
 primaryClass = {nucl-th},
       adsurl = {https://ui.adsabs.harvard.edu/abs/2001PhRvC..65a1301H}
}

@article{KALZIGITOV2020Bul,
title = "{A microscopic two-cluster model of processes in $^{6}$Li}",
journal = "News Nat. Acad. Scien. Rep. Kazakhstan: Phys.-Math. Ser. ",
volume = "4",
number = "332",
pages = "86-94",
year = "2020",
issn = "1991-346X",
doi = "https://doi.org/10.32014/2020.2518-1726.69",
author = {{Kalzhigitov}, N. and {Takibayev},  N.~Zh. and {Vasilevsky}, V.~S. and  {Akzhigitova}, E.~M. and {Kurmangaliyeva}, V.~O.}
}

@ARTICLE{2025arXiv250813702K,
       author = {{Kalzhigitov}, N.~K. and {Amangeldinova}, S. and {Vasilevsky}, V.~S.},
        title = "{Structure of hypernucleus $_{\Lambda}^{7}$Li within microscopic three-cluster model}",
      journal = {arXiv e-prints},
         year = 2025,
        month = aug,
          eid = {arXiv:2508.13702},
        pages = {arXiv:2508.13702},
archivePrefix = {arXiv},
       eprint = {2508.13702},
 primaryClass = {nucl-th},
       adsurl = {https://ui.adsabs.harvard.edu/abs/2025arXiv250813702K}
}

@BOOK{kn:kukulin89,
      AUTHOR=    "Kukulin, V.I. and Krasnopol'sky, V.M. and Hor\'{a}\v{c}ek, J.",
      TITLE=     "Theory of Resonances. Principles and Applications",
      PUBLISHER=   "Kluwer Academic Publishers",
      YEAR=      "1989",
      ADDRESS=    "Dordrecht",
      PAGES=     ""
      }
\bibliographystyle{apsrev4-2} 
\end{document}